\documentclass[a4paper, 11pt]{article}
\usepackage{amssymb}
\usepackage{amsmath}
\usepackage[all]{xy}
\usepackage{verbatim}

\newcommand{\be}{\begin{equation}}
\newcommand{\ee}{\end{equation}}
\newcommand{\bea}{\begin{eqnarray}}
\newcommand{\eea}{\end{eqnarray}}

\newcommand{\ri}{\mathrm{i}}

\newcommand{\bN}{\mathbb{N}}
\newcommand{\bR}{\mathbb{R}}
\newcommand{\bC}{\mathbb{C}}

\newcommand{\bP}{\mathbb{P}}

\newcommand{\cC}{\mathcal{C}}
\newcommand{\cD}{\mathcal{D}}

\newcommand{\cL}{\mathcal{L}}
\newcommand{\cP}{\mathcal{P}}
\newcommand{\cR}{\mathcal{R}}

\newcommand{\cO}{\mathcal{O}}
\newcommand{\cV}{\mathcal{V}}
\newcommand{\cF}{\mathcal{F}}
\newcommand{\cA}{\mathcal{A}}
\newcommand{\cM}{\mathcal{M}}
\newcommand{\cS}{\mathcal{S}}

\newcommand{\mfa}{\mathfrak{a}}
\newcommand{\mfc}{\mathfrak{c}}
\newcommand{\mfm}{\mathfrak{m}}
\newcommand{\mfu}{\mathfrak{u}}
\newcommand{\mfg}{\mathfrak{g}}
\newcommand{\mfgl}{\mathfrak{gl}}
\newcommand{\mfk}{\mathfrak{k}}
\newcommand{\mfp}{\mathfrak{p}}
\newcommand{\mfX}{\mathfrak{X}}
\newcommand{\mfs}{\mathfrak{s}}
\newcommand{\mfV}{\mathfrak{V}}
\newcommand{\mfL}{\mathfrak{L}}

\newcommand{\bsone}{\boldsymbol{1}}
\newcommand{\bsq}{\boldsymbol{q}}
\newcommand{\bsp}{\boldsymbol{p}}
\newcommand{\bslambda}{\boldsymbol{\lambda}}
\newcommand{\bschi}{\boldsymbol{\chi}}
\newcommand{\bsX}{\boldsymbol{X}}

\newcommand{\diag}{\text{diag}}
\newcommand{\offdiag}{{\text{off-diag}}}
\newcommand{\pr}{{\text{pr}}}

\newcommand{\ad}{\mathrm{ad}}

\newcommand{\Ad}{\mathrm{Ad}}
\newcommand{\tr}{\mathrm{tr}}
\newcommand{\Id}{\mathrm{Id}}

\newcommand{\reg}{\mathrm{reg}}

\newcommand{\ddd}{\mathrm{d}}
\newcommand{\ext}{\mathrm{ext}}

\newcommand{\eps}{\varepsilon}
\newcommand{\acts}{\, . \,}
\newcommand{\inner}{\, \lrcorner \,}

\newcommand{\half}{\frac{1}{2}}

\newcommand{\qedsymb}
{\hspace*{\stretch{1}}$\blacksquare$}

\renewcommand{\theequation}
{\arabic{section}.\arabic{equation}}

\begin{document}
\begin{center}
\Large{\textbf{
Action-angle duality between the $C_n$-type
hyperbolic Sutherland and the rational  
Ruijsenaars--Schneider--van Diejen models 
}}
\end{center}
\bigskip
\begin{center}
B.G.~Pusztai\\
Bolyai Institute, University of Szeged,\\
Aradi v\'ertan\'uk tere 1, H-6720 Szeged, 
Hungary\\
e-mail: \texttt{gpusztai@math.u-szeged.hu}
\end{center}
\bigskip
\begin{abstract}
In a symplectic reduction framework we construct
action-angle systems of canonical coordinates for 
both the hyperbolic Sutherland and the 
rational Ruijsenaars--Schneider--van Diejen 
integrable models associated with the $C_n$ 
root system. The presented dual reduction picture 
permits us to establish the action-angle duality
between these many-particle systems.
\end{abstract}
\newpage
\section{Introduction}
\setcounter{equation}{0}
In the theory of integrable Hamiltonian systems 
the construction of action-angle coordinates 
is of primary interest. 
The motivation comes from the fact that
in these variables the equation of motion takes 
a particularly simple linearized form, whence 
its integration is trivial. For Liouville 
integrable systems, under certain technical 
conditions, the Liouville--Arnold theorem 
\cite{Arnold} guarantees the existence of 
action-angle variables, but this result is 
of little help in the explicit construction of 
these distinguished canonical coordinates. 
The actual fact is that a bare-hand attempt 
for their construction may easily lead to 
non-trivial analytic subtleties. However, for 
certain integrable many-particle systems defined 
on the real line the construction of the 
action-angle variables is under complete control. 
Indeed, in an ingenious paper \cite{RuijCMP1988} 
Ruijsenaars has constructed action-angle coordinates 
for both the Calogero--Moser--Sutherland (CMS) 
and the Ruijsenaars--Schneider (RS) models
associated with the $A_n$ root system. One of 
the outcomes of his analysis is that the
action variables of the non-relativistic models
are the particle-positions of the relativistic
models, and vice versa. This phenomenon 
goes under the name of \emph{action-angle duality}
between the CMS and the RS models. 

Though the 
construction of the action-angle variables 
has been extended to a larger class of $A_n$-type
particle systems (see e.g. \cite{RuijRIMS2}, 
\cite{RuijRIMS3}), and the duality properties 
have been also reinterpreted in the context of 
symplectic reduction 
(see e.g. \cite{FeherKlimcik0901}, 
\cite{FeherKlimcik0906}, \cite{FeherAyadi1000}), 
to our knowledge the explicit construction 
of action-angle variables for the non-$A_n$-type 
CMS and RS models has not appeared in the 
literature before. 
The main goal of this paper is to construct
action-angle variables for both the hyperbolic 
$C_n$ Sutherland and the rational $C_n$
Ruijsenaars--Schneider--van Diejen (RSvD) 
models. 

Let us recall that the phase space 
of the hyperbolic $C_n$ Sutherland model is
\be
\cP^S 
= \{ (q_1, \ldots, q_n, p_1, \ldots, p_n) 
\in \bR^n \times \bR^n 
\, | \,
q_1 > \ldots > q_n > 0 \},
\label{cP_S}
\ee
and the dynamics is generated by the Hamiltonian
\be
H^S_{C_n} = \half \sum_{c = 1}^n p_c^2
+ \sum_{1 \leq a < b \leq n}
\left(
\frac{g^2}{\sinh^2(q_a - q_b)}
+ \frac{g^2}{\sinh^2(q_a + q_b)}
\right)
+ \half \sum_{c = 1}^n \frac{g_2^2}{\sinh^2(2 q_c)},
\label{H_Sutherland}
\ee 
where the so-called \emph{coupling parameters}
$g$ and $g_2$ are arbitrary non-zero real
numbers. As is known, the dynamics admits a 
Lax representation, which in turn provides 
a simple solution algorithm for the model. 
Also, the non-$A_n$-type Sutherland models 
have been successfully fitted into a convenient 
symplectic reduction framework 
(for details see e.g. 
\cite{OlshaPere76},
\cite{OlshaPere},
\cite{PerelomovBook},
\cite{AvanBabelonTalon1994}, 
\cite{FeherPusztai2006},
\cite{FeherPusztai2007}).
By furnishing action-angle variables for the 
hyperbolic $C_n$ Sutherland model, in this paper 
we complete their symplectic geometric 
understanding. 

The phase space of the rational $C_n$
RSvD model is
\be
\cP^R 
= \{ (\lambda_1, \ldots, \lambda_n, 
\theta_1, \ldots, \theta_n) 
\in \bR^n \times \bR^n 
\, | \,
\lambda_1 > \ldots > \lambda_n > 0 \},
\label{cP_R}
\ee
and the dynamics is governed by the Hamiltonian 
function
\be
H^R_{C_n} 
= \sum_{c = 1}^{n} \cosh(2 \theta_c)
\left( 1 
+ \frac{g_2^2}
{\lambda_c^2} \right)^\frac{1}{2}
\prod_{\substack{a = 1 \\ (a \neq c)}}^{n}
\left( 1 
+ \frac{4 g^2}
{(\lambda_c - \lambda_a)^2} \right)^\frac{1}{2}
\left( 1 
+ \frac{4 g^2}
{(\lambda_c + \lambda_a)^2} \right)^\frac{1}{2}.
\label{H_RSvD}
\ee
The geometric aspects of this model is far
less developed than that of the Sutherland
models. Although the non-$A_n$-type
RSvD models are also known to be Liouville 
integrable \cite{vanDiejen1994}, even the Lax 
representation of their dynamics is missing. 
However, in our recent paper \cite{Pusztai_JPA} 
we proposed a Lax matrix for the rational $C_n$ 
RSvD model (\ref{H_RSvD}) with two independent 
coupling parameters. In this paper we prove that 
the proposed Lax matrix (\ref{cA_def}) does have 
the properties one expects from Lax matrices. 
What is more important, we construct action-angle 
variables and also provide a solution algorithm 
for the RSvD model (\ref{H_RSvD}).
As a by-product of our construction,
we establish the aforementioned action-angle
duality between the hyperbolic $C_n$ Sutherland
and the rational $C_n$ RSvD models.

We now briefly outline the content of the
rest of the paper. Since we work out the
construction of the action-angle coordinates
in a symplectic reduction framework, section 2
is devoted to be a brief review on the reduction 
procedure. Using this machinery, in section 3 
we derive the phase space of the hyperbolic 
$C_n$ Sutherland model from symplectic reduction. 
Though this is a standard material, we find it 
instructive to discuss the reduction picture of 
the Sutherland models, mainly to see the parallel 
development with the theory of the RSvD models. 
The presentation of the new results starts with 
section 4. First we examine the main properties 
of the proposed Lax matrix (\ref{cA_def}) for 
the $C_n$ RSvD model. The main new technical
result of the paper is theorem 14, in which
we show that the parametrization of the
Lax matrix (\ref{cA_def}) does provide a 
canonical coordinate system. Having equipped 
with theorems 5 and 14, in section 5 we construct 
action-angle coordinates for 
both the $C_n$ Sutherland and the $C_n$ RSvD 
models in one stroke. By the very nature of the 
construction, the action-angle duality between 
these models comes for free. Also, the
proposed geometric picture naturally leads 
to simple solution algorithms for both models.
Incidentally, we prove the action-angle
duality between two families of Hamiltonian
systems, among which the Sutherland model
(\ref{H_Sutherland}) and the RSvD model
(\ref{H_RSvD}) are undoubtedly the physically 
most interesting members. Finally, we conclude 
the paper with an appendix on some useful 
facts on Cauchy matrices.

\section{Preliminaries}
\setcounter{equation}{0}
Take an arbitrary $n \in \bN = \{1, 2, \ldots\}$ 
and let $N = 2 n$. Making use of the 
$N \times N$ unitary matrix
\be
C = 
\begin{bmatrix}
0_n & \bsone_n \\
\bsone_n & 0_n
\end{bmatrix}
\in U(N),
\label{C}
\ee
we define the non-compact real reductive matrix 
Lie group
\be
G = U(n, n) 
= \{ y \in GL(N, \bC) \, | \, y^* C y = C \}
\label{G}
\ee
of dimension $\dim(G) = 4 n^2$. Its Lie algebra
has the form 
\be
\mfg = \mfu(n, n) 
= \{ Y \in \mfgl(N, \bC) 
\, | \, Y^* C + C Y = 0 \}.
\label{mfg}
\ee
Notice that the map
\be
\langle \, , \rangle \colon 
\mfg \times \mfg \rightarrow \bR,
\quad
(Y_1, Y_2) \mapsto 
\langle Y_1, Y_2 \rangle = \tr(Y_1 Y_2)
\label{bilinear_form}
\ee
is a well-defined $\Ad$-invariant symmetric 
bilinear form on $\mfg$. Since it is also 
non-degenerate, the map
\be
\mfg \ni Y \mapsto \langle Y, \cdot \rangle 
\in \mfg^*
\label{mfg_and_its_dual}
\ee
is a linear isomorphism, leading to the natural 
identification $\mfg^* \cong \mfg$. Moreover, 
by its $\Ad$-invariance, the bilinear form 
(\ref{bilinear_form}) naturally induces a 
biinvariant pseudo-Riemannian metric on the 
group $G$. 
Now let $L_a \colon y \mapsto a y$ $(a \in G)$ 
denote the left translations 
on $G$; then the left trivialization
\be
G \times \mfg^* \ni (y, \varphi) 
\mapsto (L_{y^{-1}})^* \varphi \in T^* G
\label{cot_bundle_identification} 
\ee
permits us to identify the cotangent bundle 
$T^* G$  with $G \times \mfg^*$. Moreover, 
due to the isomorphism (\ref{mfg_and_its_dual}), 
the product manifold
\be
\cP = G \times \mfg
\label{P}
\ee
is also diffeomorphic to $T^* G$. For 
convenience, in the rest of the paper we 
use $\cP$ as an appropriate model of the 
cotangent bundle of $G$. Observe that the 
tangent spaces of $\cP$ can be identified as
\be
T_{(y, Y)} \cP = T_{(y, Y)}(G \times \mfg) 
\cong T_y G \oplus T_Y \mfg 
\cong T_y G \oplus \mfg
\qquad
((y, Y) \in \cP).
\label{tangent_space_identification}
\ee
Regarding the symplectic structure of 
$\cP \cong T^* G$, the canonical $1$-form 
$\vartheta \in \Omega^1(\cP)$ can be written 
as
\be
\vartheta_{(y, Y)} (\delta y \oplus \delta Y) 
= \langle y^{-1} \delta y, Y \rangle,
\label{canonical_theta}
\ee
and for the canonical symplectic form 
$\omega = -\ddd \vartheta \in \Omega^2(\cP)$ 
we have 
\be
\omega_{(y, Y)}(\Delta y \oplus \Delta Y, 
\delta y \oplus \delta Y)
= \langle y^{-1} \Delta y, \delta Y \rangle
- \langle y^{-1} \delta y, \Delta Y \rangle
+ \langle [y^{-1} \Delta y, y^{-1} \delta y], 
Y \rangle,
\label{omega}
\ee
where $(y, Y) \in \cP$ is an arbitrary point 
and $\Delta y \oplus \Delta Y, 
\delta y \oplus \delta Y \in T_y G \oplus \mfg$ 
are arbitrary tangent vectors. (For more details 
see e.g. Proposition 4.4.2 in \cite{AM}.)

Now, let us consider the Cartan involution 
$\Theta(y) = (y^{-1})^*$ $(y \in G)$ and its 
fixed-point set
\be
K = \{ y \in G \, | \, \Theta(y) = y \}
= \{ y \in G \, | \, y \mbox{ is unitary}\} 
\cong U(n) \times U(n).
\ee
Note that $K$ is a maximal compact subgroup 
of $G$ of dimension $\dim(K) = 2 n^2$, 
in which the matrix $C$ (\ref{C}) is a 
central element. The Lie algebra involution 
$\theta(Y) = -Y^*$ $(Y \in \mfg)$ corresponding 
to $\Theta$ induces the Cartan decomposition
\be
\mfg = \mfk \oplus \mfp
\label{gradation}
\ee
with the Lie subalgebra and the complementary 
subspace
\be
\mfk = \ker(\theta - \Id) 
= \{ Y \in \mfg \, | \, Y^* = - Y \}
\quad \mbox{and} \quad
\mfp = \ker(\theta + \Id) 
= \{ Y \in \mfg \, | \, Y^* = Y \},
\label{mfkp}
\ee
respectively. That is, any element $Y \in \mfg$ 
can be uniquely decomposed as
\be
Y = Y_+ + Y_- 
\qquad
(Y_+ \in \mfk, \, Y_- \in \mfp).
\label{decomp}
\ee
Notice that $\dim(\mfk) = \dim(\mfp) = 2 n^2$.
We mention in passing that the bilinear form 
(\ref{bilinear_form}) is negative definite
(respectively positive definite) on 
$\mfk$ (respectively on $\mfp$).

Next, with the aid of $K$ we introduce 
an isometric group action on $G$. Indeed, 
the formula 
\be
(k_L, k_R) \acts y = k_L y k_R^{-1}
\qquad
(y \in G, \, (k_L, k_R) \in K \times K)
\label{KKonG}
\ee
defines a smooth left action of the product 
Lie group $K \times K$ on $G$. The natural 
lift of this action onto $\cP \cong T^* G$ 
has the form 
$\Phi \colon 
(K \times K) \times \cP \rightarrow \cP$ 
with
\be
\Phi_{(k_L, k_R)}(y, Y) 
= (k_L, k_R) \acts (y, Y) 
= (k_L y k_R^{-1}, k_R Y k_R^{-1})
\qquad
((y, Y) \in \cP, 
\, (k_L, k_R) \in K \times K).
\label{KKonP}
\ee
For any $X_L \oplus X_R \in \mfk \oplus \mfk$ 
let the vector field 
$(X_L \oplus X_R)^\sharp \in \mfX(\cP)$ denote 
the corresponding infinitesimal generator of 
$\Phi$. Obviously at each point $(y, Y) \in \cP$ 
we have
\be
(X_L \oplus X_R)^\sharp_{(y, Y)} 
= (X_L y - y X_R) \oplus [X_R, Y] 
\in T_y G \oplus \mfg.
\ee
Since action $\Phi$ leaves the canonical 
$1$-form (\ref{canonical_theta}) invariant, 
i.e. for every group element
$(k_L, k_R) \in K \times K$ we have 
$\Phi_{(k_L, k_R)}^* \vartheta = \vartheta$,
the corresponding momentum map 
$J \colon \cP \rightarrow (\mfk \oplus \mfk)^*$ 
takes the form
\be
J(y, Y)(X_L \oplus X_R) 
= \vartheta_{(y, Y)} 
\left( (X_L \oplus X_R)^\sharp_{(y, Y)} \right)  
= \langle (y Y y^{-1})_+, X_L \rangle 
+ \langle -Y_+, X_R \rangle.
\ee
(See e.g. Theorem 4.2.10 in \cite{AM}.)
Now note that the formula
\be
\langle Z_L \oplus Z_R, 
X_L \oplus X_R \rangle_{\mfk \oplus \mfk}
= \langle Z_L, X_L \rangle 
+ \langle Z_R, X_R \rangle
\label{scalar_prod_on_kk}
\ee
defines an invariant negative definite 
symmetric bilinear form on $\mfk \oplus \mfk$, 
whence the linear isomorphism
\be
\mfk \oplus \mfk \ni Z_L \oplus Z_R 
\mapsto \langle Z_L \oplus Z_R , 
\cdot \rangle_{\mfk \oplus \mfk}
\in (\mfk \oplus \mfk)^*
\label{kk_dual_identification}
\ee
allows us to make the natural identification 
$(\mfk \oplus \mfk)^* \cong \mfk \oplus \mfk$.
Therefore the momentum map for action $\Phi$ 
can be realized as a $K \times K$-equivariant 
map of the form
\be
J \colon \cP \rightarrow \mfk \oplus \mfk,
\quad (y, Y) \mapsto J(y, Y) 
= (y Y y^{-1})_+ \oplus (-Y_+).
\label{J}
\ee

Now, with any column vector $V \in \bC^N$ 
satisfying the conditions $V^* V = N$ and 
$C V + V = 0$ we associate the traceless 
$N \times N$ matrix
\be
\xi(V) 
= \ri g (V V^* - \bsone_N) + \ri (g - g_2) C 
\in \mfk.
\label{xi}
\ee
Also, let $E \in \bC^N$ denote the distinguished 
column vector with components
\be
E_a = - E_{n + a} = 1 \qquad 
(a \in \bN_n = \{1, \ldots, n \}).
\ee
In order to understand the action-angle 
duality between the hyperbolic  $C_n$ Sutherland 
and the rational $C_n$ RSvD models from a 
symplectic reduction picture, we wish to reduce
the Hamiltonian $K \times K$-space 
$(\cP, \omega, \Phi, J)$ at the very special 
value\footnote{Note that by fixing $J$ to the 
slightly more general value 
$-\xi(E) \oplus \ri \kappa C \in \mfk \oplus \mfk$ 
depending on a new real parameter $\kappa \in \bR$, 
we can even understand the action-angle duality 
between the hyperbolic $BC_n$ Sutherland and the 
rational $BC_n$ RSvD models with \emph{three 
independent coupling parameters}. By working out the 
$C_n$ case in this paper, we also pave the way for 
the analysis of the more involved $BC_n$ case, 
which we wish to publish elsewhere. 
}
\be
\mu = - \xi(E) \oplus 0 
\in \mfk \oplus \mfk \cong (\mfk \oplus \mfk)^*
\label{mu}
\ee
of the momentum map $J$. To perform the 
actual reduction, we utilize the 
so-called shifting trick (see e.g. \cite{OR}).
Recall that the shifting trick relies on 
the co-adjoint orbit passing through $-\mu$,
which in our case can be identified with 
the adjoint orbit
$\cO \oplus \{ 0 \} \subset \mfk \oplus \mfk 
\cong (\mfk \oplus \mfk)^*$, where
\be
\cO = \cO(\xi(E)) 
= \{ \xi(V) \in \mfk \, | \, V \in \bC^N, \,
V^* V = N, \, C V + V = 0 \}.
\label{cO}
\ee
Since $\cO \cong \bC \bP ^{n - 1}$, for its real 
manifold dimension we have $\dim(\cO) = 2 n - 2$. 
Also, the tangent spaces of $\cO$ can be 
identified as
\be
T_\rho \cO \cong \{ \ad_X (\rho) = [X, \rho] 
\, | \, 
X \in \mfk \} \subset \mfk
\qquad
(\rho \in \cO),
\label{cO_tangent_space}
\ee
and the natural Kirillov--Kostant--Souriau 
symplectic form $\omega^\cO \in \Omega^2(\cO)$ 
carried by $\cO$ has the form
\be
\omega^\cO_\rho( \ad_X (\rho), \ad_Z (\rho) ) 
= \langle \rho, [X, Z] \rangle
\qquad
(\rho \in \cO, \, X, Z \in \mfk).
\label{omega_cO}
\ee
Following the idea of the shifting trick, 
we enlarge the initial phase space $\cP$ by the 
orbit $\cO \oplus \{ 0 \} \cong \cO$, i.e. we 
consider the extended phase space
\be
\cP^\ext = \cP \times \cO 
= \{ (y, Y, \rho) 
\, | \, 
y \in G, \, Y \in \mfg, \, \rho \in \cO \}, 
\label{Pext}
\ee
and endow it with the product symplectic 
structure 
\be
\omega^\ext = \omega + \omega^\cO.
\label{omega_ext}
\ee
The natural diagonal action of  
$K \times K$ on $\cP^\ext$ is given by 
$\Phi^\ext \colon 
(K \times K) \times \cP^\ext 
\rightarrow \cP^\ext$,
where
\be
\Phi^\ext_{(k_L, k_R)}(y, Y, \rho) 
= (k_L, k_R) \acts (y, Y, \rho) 
= (k_L y k_R^{-1}, k_R Y k_R^{-1}, 
k_L \rho k_L^{-1}),
\label{Phi_ext}
\ee
and the corresponding $K \times K$-equivariant 
momentum map 
$J^\ext \colon 
\cP^\ext \rightarrow \mfk \oplus \mfk$ 
takes the form
\be
J^\ext(y, Y, \rho) 
= ( (y Y y^{-1})_+ + \rho ) \oplus (- Y_+).
\label{Jext}
\ee
Notice, however, that 
$\tr( (y Y y^{-1})_+ + \rho ) + \tr(- Y_+) = 0$.
Therefore, upon introducing the Lie subalgebra
\be
\mfs(\mfk \oplus \mfk) 
= \{ X_L \oplus X_R \in \mfk \oplus \mfk 
\, | \, \tr(X_L) + \tr(X_R) = 0 \}
\leq \mfk \oplus \mfk,
\ee
we can write 
$J^\ext (y, Y, \rho) \in \mfs(\mfk \oplus \mfk)$
$(\forall (y, Y, \rho) \in \cP^\ext)$; that is, 
$J^\ext$ is actually an 
$\mfs(\mfk \oplus \mfk)$-valued map.  

The shifting trick, which basically states 
that the reduced phase space 
$\cP /\!/_{\mu} (K \times K)$ is 
symplectomorphic to the symplectic quotient 
$\cP^\ext /\!/_0 (K \times K)$, is only a 
convenient technical tool in our present work. 
In general this approach does not make the 
reduction simpler, only relegates the 
differential-topological difficulties into 
the orbit part of $\cP^\ext$. However, our 
experience with the 
Calogero--Moser--Sutherland-type 
many-particle systems convinces us 
that the shifting trick does provide a cleaner 
and shorter derivation of these integrable systems
in the symplectic reduction framework 
(see e.g \cite{FeherPusztai2006}, 
\cite{FeherPusztai2007}).

Finally, recall that the Hamiltonian vector 
field $\bsX_H \in \mfX(\cP^\ext)$ associated 
with an arbitrary smooth function 
$H \in C^\infty(\cP^\ext)$ is (uniquely)
determined by the condition
\be
\ddd H = \bsX_H \inner \omega^\ext,
\label{X_H}
\ee
and the Poisson bracket of any pair of functions 
$F, H \in C^\infty(\cP^\ext)$ is defined as
\be
\{ F, H \}^\ext = \omega^\ext(\bsX_F, \bsX_H)
= \bsX_H [F].
\label{PB}
\ee

\section{The phase space of the hyperbolic 
$C_n$ Sutherland model}
\setcounter{equation}{0}
The first step of the Marsden--Weinstein 
reduction is to solve the momentum map
constraint 
\be
J^\ext(y, Y, \rho) = 0
\label{constraint}
\ee
for $(y, Y, \rho) \in \cP^\ext 
= G \times \mfg \times \cO$; that is, we
have to understand the properties of the
level set  
\be
\mfL_0 = (J^\ext)^{-1}(\{ 0 \}) \subset \cP^\ext.
\label{cL_0}
\ee 
Our first goal in this section is to solve the 
constraint (\ref{constraint}) by diagonalizing 
the group elements $y \in G$. However, to make 
the notion of diagonalization precise, we need 
some more Lie theoretic facts, which we summarize
in the following subsection. For a general 
reference on the related group theoretic issues
we recommend \cite{Knapp}.

\subsection{Group theoretic background}
For any real $n$-tuple 
$q = (q_1, \ldots, q_n) \in \bR^n$ we define 
$\bsq = \diag(q_1, \ldots, q_n)$ and
$Q = \diag(\bsq, - \bsq) \in \mfp$. Clearly 
the set of diagonal matrices
\be
\mfa = \{ Q \in \mfp \, | \, q \in \bR^n \}
\ee
forms a maximal Abelian subspace in $\mfp$.
If $\mfa^\perp$ denotes the set of the 
off-diagonal elements of $\mfp$, then we have 
the orthogonal decomposition 
$\mfp = \mfa \oplus \mfa^\perp$.
Note that the centralizer of $\mfa$ inside $K$ 
is the Abelian Lie group
\be
M = Z_K(\mfa) 
= \{ \diag( e^{\ri \bschi}, e^{\ri \bschi} ) 
\in K \, | \, \chi \in \bR^n \}
\leq K
\label{M}
\ee
with Lie algebra $\mfm 
= \{ \diag( \ri \bschi, \ri \bschi ) 
\in \mfk \, | \, \chi \in \bR^n \} \leq \mfk$.
If $\mfm^\perp$ denotes the set of 
the off-diagonal elements of $\mfk$, we can 
write $\mfk = \mfm \oplus \mfm^\perp$. Since 
the subspace $\mfm^\perp \oplus \mfa^\perp$
consists of the off-diagonal elements of $\mfg$,
the restricted operator
\be
\tilde{\ad}_Q 
= \ad_Q |_{\mfm^\perp \oplus \mfa^\perp}
\in \mfgl(\mfm^\perp \oplus \mfa^\perp)
\label{tilde_ad}
\ee
is well-defined for any $q \in \bR^n$, and its 
spectrum has the form
\be
\sigma(\tilde{\ad}_Q) 
= \{q_a - q_b, \pm (q_a + q_b), \pm 2 q_c 
\, | \, 
a, b, c \in \bN_n, \, a \neq b\}.
\ee
The regular part of $\mfa$ is defined as 
$\mfa_\reg = \{ Q \in \mfa 
\, | \, 
\tilde{\ad}_Q \mbox{ is invertible} \}$,
and the standard Weyl chamber
\be
\mfc = \{ Q \in \mfa_\reg 
\, | \, q_1 > \ldots > q_n > 0 \}
\label{mfc}
\ee
is an appropriate connected component of 
$\mfa_\reg$. Without any further notice, 
in the rest of the paper we frequently 
identify $\mfc$ with the subset 
$\{ q = (q_1, \ldots, q_n) \in \bR^n 
\, | \, q_1 > \ldots > q_n > 0 \} 
\subset \bR^n$.

At the Lie algebra level it is a crucial fact 
that the elements of  
$\mfp$ (\ref{mfkp}) can be conjugated 
into $\mfa$ by appropriate elements of $K$. 
More precisely, the map
\be
\mfa \times K \ni (Q, k) 
\mapsto k Q k^{-1} \in \mfp
\label{diagonalizing_mfp}
\ee
is well-defined and onto. The regular part
of $\mfp$ defined by 
$\mfp_\reg = \{ k Q k^{-1} 
\, | \, Q \in \mfc, k \in K \}$
is dense and open in $\mfp$. Moreover, 
the map
\be
\mfc \times K / M \ni (Q, k M) 
\mapsto k Q k^{-1} \in \mfp_\reg
\label{mfp_reg_parametrization}
\ee
is a diffeomorpism, allowing us to make the 
identification 
$\mfp_\reg \cong \mfc \times K / M$.
 
To proceed further we need some basic facts from 
the canonical forms of the Lie group elements, 
too. First, recall that the global Cartan 
decomposition (polar decomposition) 
\be
\exp(\mfp) \times K \ni (e^Y, k) 
\mapsto e^Y k \in G
\label{global_Cartan_decomposition}
\ee
is a diffeomorphism. Combining this fact with 
(\ref{diagonalizing_mfp}), it is obvious that 
any group element $y \in G$ can be written as 
\be
y = k_L e^Q k_R^{-1} 
\label{KAK}
\ee
with some $k_L, k_R \in K$ and $Q \in \mfa$. 
Upon setting $A = \exp(\mfa)$, 
the factorization (\ref{KAK}) of the elements of 
$G$ is usually called the $K A K$ decomposition. 

Let us now define the regular part of the Abelian
Lie group $A$ by 
$A_\reg = \{ e^Q \in A \, | \, Q \in \mfa_\reg \}$,
and also let $G_\reg = K A_\reg K$. It is known 
that $G_\reg$ is a dense and open submanifold of 
$G$. Moreover, if $M_*$ denotes the diagonal 
embedding of $M$ (\ref{M}) into $K \times K$, i.e
$M_* 
= \{ (m, m) \in K \times K \, | \, m \in M \}$,
then the map
\be
\mfc \times (K \times K) / M_* \ni 
(Q, (k_L, k_R) M_*) \mapsto k_L e^Q k_R^{-1} 
\in G_\reg
\ee
is a diffeomorphism. Thus the identification
$G_\reg \cong \mfc \times (K \times K) / M_*$
is immediate.

\subsection{Parametrization of $\mfL_0$ induced
by the $KAK$ decomposition}
Having equipped with the necessary group 
theoretic facts, now we are in a position to 
derive both the hyperbolic $C_n$ Sutherland 
and the rational $C_n$ RSvD models from a 
uniform symplectic reduction framework.
To derive the Sutherland system 
(\ref{H_Sutherland}) defined on  
(\ref{cP_S}), one of  the key ingredients 
is the Lax operator
\be
L \colon \cP^S
\rightarrow \mfp,
\quad
(q, p) \mapsto L(q, p) =
P - \sinh(\tilde{\ad}_Q)^{-1} \xi(E),
\label{L}
\ee 
where we employ the notation 
$P = \diag(\bsp, -\bsp) \in \mfa$ with 
$\bsp = \diag(p_1, \ldots, p_n)$, 
as we stipulated in the previous
subsection. Notice that the matrix entries 
of $L$ have the form
\begin{align}
& L_{a, b} = - L_{n + a, n + b} 
= \frac{-\ri g}{\sinh(q_a - q_b)},
& & L_{a, n + b} = - L_{n + a, b} 
= \frac{\ri g}{\sinh(q_a + q_b)},
\\
& L_{c, n + c} = - L_{n + c, c}
= \frac{\ri g_2}{\sinh(2 q_c)},
& & L_{c, c} = - L_{n + c, n + c} 
= p_c,
\label{L_entries} 
\end{align}
where $a, b, c \in \bN_n$ and $a \neq b$.
Now, making use of the Lax operator $L$, 
the $K A K$ decomposition of $G$ leads to 
the following characterization of the points 
of the level set $\mfL_0$ (\ref{cL_0}).

\smallskip
\noindent
\textbf{Proposition 1.}
\emph{
For each  
$(y, Y, \rho) \in \mfL_0$
there exist some $q \in \mfc$, 
$p \in \bR^n$ and $\eta_L, \eta_R \in K$ such 
that
\be
y = \eta_L e^Q \eta_R^{-1},
\quad
Y = \eta_R L(q, p) \eta_R^{-1},
\quad
\rho = \eta_L \xi(E) \eta_L^{-1}.
\label{proposition_1}
\ee
}

\smallskip
\noindent
\textbf{Proof.}
Take an arbitrary 
$(y, Y, \rho) \in \mfL_0$;
then by the $K A K$ decomposition (\ref{KAK}) 
we have the factorization 
\be
y = \tilde{\eta}_L e^Q \tilde{\eta}_R^{-1} 
\ee
with some 
$\tilde{\eta}_L, \tilde{\eta}_R \in K$
and $Q = \diag(\bsq, -\bsq) \in \mfa$
satisfying $q_1 \geq \ldots \geq q_n \geq 0$.
Also, due to (\ref{cO}), we can write 
$\rho = \xi(V)$ with some $V \in \bC^N$ 
satisfying $V^* V = N$ and $C V + V = 0$. 
Plugging this parametrization into $J^\ext$
(\ref{Jext}), we conclude that $Y_+ = 0$, 
i.e. $Y = Y_- \in \mfp$, together with
\be
0 = (y Y y^{-1})_+ + \xi(V)
= \tilde{\eta}_L \left( 
\sinh(\ad_Q) 
(\tilde{\eta}_R^{-1} Y_- \tilde{\eta}_R) 
+ \xi(\tilde{\eta}_L^{-1} V) 
\right) \tilde{\eta}_L^{-1}.
\ee
Upon introducing 
$\tilde{Y} 
= \tilde{\eta}_R^{-1} Y_- \tilde{\eta}_R 
\in \mfp$ 
and 
$\tilde{V} = \tilde{\eta}_L^{-1} V \in \bC^N$, 
the above equation can be rewritten as
\be
\sinh(\ad_Q) \tilde{Y} = -\xi(\tilde{V}).
\label{tilde_Y}
\ee
Componentwise, for any $k, l \in \bN_N$ 
we have
\be
\sinh(q_k - q_l) \tilde{Y}_{k, l} 
= \ri g (\delta_{k, l} 
- \tilde{V}_k \overline{\tilde{V}}_l) 
+ \ri (g_2 - g) C_{k, l},
\label{tilde_Y_entries}
\ee
where it is understood that $q_{n + a} = - q_a$
$(\forall a \in \bN_n)$.

Take an arbitrary $a \in \bN_n$; then with  
$k = l = a$ we get
$0 =  \ri g (1 - \vert \tilde{V}_a \vert^2)$,
therefore 
$\tilde{V}_a = - \tilde{V}_{n + a} 
= e^{\ri \chi_a}$
with some $\chi_a \in \bR$. With $k = a$ and 
$l = n + a$ we obtain 
$\sinh(2 q_a) \tilde{Y}_{a, n + a} 
= \ri g_2 \neq 0$, 
whence $q_a \neq 0$ also follows.

Next, let $a, b \in \bN_n$, $a \neq b$;
then with $k = a$ and $l = b$ we have
$\sinh(q_a - q_b) \tilde{Y}_{a, b}
= - \ri g \tilde{V}_a \overline{\tilde{V}}_b
\neq 0$, which implies $q_a \neq q_b$.
At this point we see that actually $q \in \mfc$.
Moreover, upon introducing the group element
$m = \diag(e^{\ri \bschi}, e^{\ri \bschi}) 
\in M$, the application of the linear operator 
$\Ad_{m^{-1}}$ on (\ref{tilde_Y}) yields
\be
\sinh(\ad_Q) \Ad_{m^{-1}} \tilde{Y}
= - \xi(E) \in \mfm^\perp,
\ee
therefore $\Ad_{m^{-1}} \tilde{Y}
= P - \sinh(\tilde{\ad}_Q)^{-1} \xi(E) = L(q, p)$
with some $p \in \bR^n$. Finally, by letting 
$\eta_L = \tilde{\eta}_L m$ 
and $\eta_R = \tilde{\eta}_R m$, the 
parametrization (\ref{proposition_1}) immediately 
follows. 
\qedsymb

The characterization of the points of the level
set $\mfL_0$ given in proposition 1
permits us to introduce a parametrization 
naturally induced by the $K A K$ decomposition 
of the group elements. For, we need the diagonal 
embedding of $U(1)$ into $K \times K$; i.e. 
the Lie subgroup
\be
U(1)_* 
= \{ (e^{\ri \chi} \bsone_N, e^{\ri \chi} \bsone_N) 
\in K \times K \, | \, 
\chi \in \bR \} \cong U(1), 
\ee 
and the corresponding Lie algebra
\be
\mfu(1)_* 
= \{ (\ri \chi \bsone_N, \ri \chi \bsone_N) 
\in \mfk \oplus \mfk \, | \, \chi \in \bR \}
\cong \mfu(1).
\ee
Also, let us define the smooth product manifold
\be
\cM^S 
= \cP^S \times (K \times K) / U(1)_*.
\label{cM}
\ee
Since $U(1)_*$ is a normal subgroup 
of $K \times K$, the coset space 
$(K \times K) / U(1)_*$ is a (real)
Lie group in a natural manner of dimension 
$4 n^2 - 1$. Note also that 
$\dim(\cM^S) = 4 n^2 + 2 n - 1$. 

\smallskip
\noindent
\textbf{Lemma 2.}
\emph{
The map 
\be
\Upsilon^S \colon  
\cM^S \rightarrow \cP^\ext, 
\quad
(q, p, (\eta_L, \eta_R) U(1)_*)  
\mapsto (\eta_L e^Q \eta_R^{-1},
\eta_R L(q, p) \eta_R^{-1}, 
\eta_L \xi(E) \eta_L^{-1})
\label{Upsilon_S}
\ee
is an injective immersion with image 
$\Upsilon^S(\cM^S) = \mfL_0$.
}

\smallskip
\noindent
\textbf{Proof.}
It is obvious that $\Upsilon^S$ is a well-defined 
smooth map. Our next goal is to determine the
image of $\cM^S$ under the map $\Upsilon^S$. 
To this end let  $q \in \mfc$, 
$p \in \bR^n$ and  $\eta_L, \eta_R \in K$ be 
arbitrary elements and consider the matrices
\be
y = \eta_L e^Q \eta_R^{-1} \in G, 
\quad
Y = \eta_R L(q, p) \eta_R^{-1} \in \mfp, 
\quad
\rho = \eta_L \xi(E) \eta_L^{-1} \in \cO.
\ee
Notice that the relationship 
\be
y Y y^{-1} 
= \eta_L \left( \cosh(\ad_Q) L(q, p) 
+ \sinh(\ad_Q) L(q, p) \right) \eta_L^{-1}
\ee
together with (\ref{L}) immediately yields
\be
(y Y y^{-1})_+ + \rho  
= \eta_L ( \sinh(\ad_Q) L(q, p)) \eta_L^{-1} 
+ \eta_L \xi(E) \eta_L^{-1} = 0.
\ee
Since the equation $Y_+ = 0$ also holds, from 
the definition of the momentum map $J^\ext$
(\ref{Jext}) we see at once that 
$J^\ext(y, Y, \rho) = 0$.
In other words, we have the inclusion relation
$\Upsilon^S (\cM^S) \subset \mfL_0$.
Recalling proposition 1 we conclude 
that $\Upsilon^S (\cM^S) = \mfL_0$. 

To show that $\Upsilon^S$ is injective, suppose 
that
\be
\Upsilon^S 
\left( q, p, (\eta_L, \eta_R) U(1)_* \right)
= \Upsilon^S \left(\tilde{q}, \tilde{p}, 
(\tilde{\eta}_L, \tilde{\eta}_R) U(1)_* \right)
\label{Upsilon_injective}
\ee
with some 
$q, \tilde{q} \in \mfc$, 
$p, \tilde{p} \in \bR^n$ and 
$\eta_L, \tilde{\eta}_L, 
\eta_R, \tilde{\eta}_R \in K$.
The $G$-component of the above equation reads as
\be
(\eta_L e^Q \eta_L^{-1}) (\eta_L \eta_R^{-1})
= \eta_L e^Q \eta_R^{-1}
= \tilde{\eta}_L e^{\tilde{Q}} \tilde{\eta}_L^{-1}
= (\tilde{\eta}_L e^{\tilde{Q}} 
\tilde{\eta}_L^{-1}) 
(\tilde{\eta}_L \tilde{\eta}_R^{-1}).
\ee 
Remembering the uniqueness of the 
global Cartan decomposition 
(\ref{global_Cartan_decomposition}), 
the relations
\be 
\eta_L e^Q \eta_L^{-1} 
= \tilde{\eta}_L e^{\tilde{Q}} 
\tilde{\eta}_L^{-1}
\quad \text{and} \quad 
\eta_L \eta_R^{-1} 
= \tilde{\eta}_L \tilde{\eta}_R^{-1}
\ee
are immediate.  
Moreover, from the regularity assumption 
$q, \tilde{q} \in \mfc$ and from the 
identification (\ref{mfp_reg_parametrization})
we see that $q = \tilde{q}$ and 
$\tilde{\eta}_L = \eta_L m$ 
with some $m \in M$. Notice that the 
relationship $\tilde{\eta}_R = \eta_R m$ 
also follows.

The $\cO$-component of (\ref{Upsilon_injective})
has the form 
\be
\eta_L \xi(E) \eta_L^{-1}
= \tilde{\eta}_L \xi(E) \tilde{\eta}_L^{-1},
\ee
hence $\xi(E) = m \xi(E) m^{-1} = \xi(m E)$. By 
(\ref{xi}) we can write 
$E E^* = (m E) (m E)^*$, which immediately 
implies that  
$m = e^{\ri \chi} \bsone_N$ 
with some $\chi \in \bR$. Therefore
$(\eta_L, \eta_R) U(1)_*
= (\tilde{\eta}_L, \tilde{\eta}_R) U(1)_*$.
 
Finally, the $\mfg$-component of 
(\ref{Upsilon_injective}) spells out as
\be
\eta_R L(q, p) \eta_R^{-1}
= \tilde{\eta}_R L(\tilde{q},\tilde{p}) 
\tilde{\eta}_R^{-1}.
\ee
Utilizing the above observations, this equation
leads to $L(q, p) = L(q, \tilde{p})$. By simply 
taking its diagonal part, we obtain 
$P = \tilde{P}$, i.e. $p = \tilde{p}$, completing 
the verification of the injectivity of 
$\Upsilon^S$.

Next, we show that  
$\Upsilon^S$ is an immersion, i.e. its derivative 
is injective at each point of $\cM^S$. For, take 
an arbitrary point 
$x = (q, p, (\eta_L, \eta_R) U(1)_*) 
\in \cM^S$. Due to the natural identifications
\be
T_q \mfc \cong \bR^n, 
\quad
T_p \bR^n \cong \bR^n,
\quad
T_e((K \times K) / U(1)_*)
\cong (\mfk \oplus \mfk) / \mfu(1)_*,
\ee
a generic tangent vector $v$ at the point $x$
has the form
\be
v = \delta q \oplus \delta p \oplus 
(\eta_L, \eta_R) (X_L \oplus X_R) \mfu(1)_* 
\in T_x \cM^S
\ee
with some $\delta q, \delta p \in \bR^n$
and $X_L, X_R \in \mfk$.
Notice also that the action of the derivative
\be
(\ddd \Upsilon^S)_x \colon 
T_x \cM^S \rightarrow T_{\Upsilon^S(x)} \cP^\ext
\ee
on $v$ is given by the formula
\be
(\ddd \Upsilon^S)_x v 
= \eta_L (e^Q \delta Q + X_L e^Q - e^Q X_R) 
\eta_R^{-1}
\oplus 
\eta_R (\delta L + [X_R, L(q, p)]) \eta_R^{-1}
\oplus
\eta_L [X_L, \xi(E)] \eta_L^{-1},
\label{derivative_Upsilon}
\ee
where $\delta L = \delta P 
+ [ \delta Q, 
\cosh(\tilde{\ad}_Q) 
\sinh(\tilde{\ad}_Q)^{-2} \xi(E)]$.

Now, to find the kernel of the 
linear map $(\ddd \Upsilon^S)_x$, suppose that
$v \in \ker((\ddd \Upsilon^S)_x)$, i.e. 
$(\ddd \Upsilon^S)_x v = 0$.  Taking
the $G$-component of $(\ddd \Upsilon^S)_x v$,
for any $k, l \in \bN_N$, $k \neq l$, we 
obtain
\be 
(X_L)_{k, l} e^{q_l} = e^{q_k} (X_R)_{k, l},
\label{XLR}
\ee
with the convention $q_{n + a} = - q_a$ 
$(\forall a \in \bN_n)$.
Combining (\ref{XLR}) with $X_L^* + X_L = 0$ and
$X_R^* + X_R = 0$, it follows that
\be
\sinh(q_k - q_l) (X_R)_{k, l} = 0.
\ee 
Since $q$ 
is regular, we see that both $X_L$ and $X_R$
are diagonal matrices, i.e. $X_L, X_R \in \mfm$.
Plugging this observation back into the 
$G$-component of $(\ddd \Upsilon^S)_x v$, we 
obtain $\delta Q + X_L - X_R = 0$. Inspecting 
its real part we get $\delta Q = 0$, i.e. 
$\delta q = 0$, meanwhile the imaginary part
yields $X_L = X_R$.

Remembering the defining formula of $\xi(E)$ 
(\ref{xi}), from the vanishing of the $\cO$-component 
of the derivative (\ref{derivative_Upsilon}) we see 
that
\be
0 = [X_L, \xi(E)] = \ri g [X_L, E E^*]
= \ri g \left( (X_L E) E^* + E (X_L E)^* \right).
\ee
Therefore, for any $a, b \in \bN_n$ 
we can write
\be
0 = (X_L E)_a \overline{E}_b 
+ E_a \overline{(X_L E)}_b
= (X_L)_{a, a} + \overline{(X_L)}_{b, b}.
\ee
It follows that
$X_L = X_R = \ri \chi \bsone_N$ with some 
$\chi \in \bR$, i.e. 
$X_L \oplus X_R \in \mfu(1)_*$. 
Finally, by inspecting the $\mfg$-component 
of (\ref{derivative_Upsilon}), we obtain
$\delta P = 0$, i.e. $\delta p = 0$.
Thus $v = 0$, meaning that the kernel of the 
derivative operator is trivial. 
\qedsymb

\subsection{Identification of the reduced 
phase space}
In this subsection we wish to 
identify the phase space of the hyperbolic $C_n$
Sutherland model in the proposed symplectic 
reduction picture. As a first step, we 
examine the differential geometric properties 
of the closed level set $\mfL_0$ (\ref{cL_0}). 

\smallskip
\noindent
\textbf{Lemma 3.}
\emph{
The zero element of $\mfs(\mfk \oplus \mfk)$ 
is a regular value of the momentum map
$J^\ext \colon 
\cP^\ext \rightarrow \mfs(\mfk \oplus \mfk)$.
} 
 
\smallskip
\noindent
\textbf{Proof.}
Pick an arbitrary 
$x = (y, Y, \rho) \in \cP^\ext$.
Working out the derivative operator 
\be 
(\ddd J^\ext)_{x} \colon
T_{x} \cP^\ext 
\rightarrow 
T_{J^\ext (x)}\mfs(\mfk \oplus \mfk)
\cong \mfs(\mfk \oplus \mfk)
\label{Jext_derivative_map}
\ee
of the momentum map $J^\ext$ (\ref{Jext}) at 
point $x$, for each tangent vector 
$\delta y \oplus \delta Y \oplus \delta \rho 
\in T_{x} \cP^\ext$ we find the formula
\be
(\ddd J^\ext)_{x} 
(\delta y \oplus \delta Y \oplus \delta \rho)
= \left(
(y (\delta Y - [Y, y^{-1} \delta y]) y^{-1})_+ 
+ \delta \rho
\right) \oplus \left( - (\delta Y)_+ \right).
\label{Jext_derivative}
\ee
Now, we have to verify that
the derivative (\ref{Jext_derivative_map}) 
is onto at each point $x \in \mfL_0$. Since 
$J^\ext$ is equivariant, by lemma 2 it  
suffices to show that (\ref{Jext_derivative_map}) 
is onto at each 
$x = (e^Q, L(q, p), \xi(E))$, 
where $q \in \mfc$ and $p \in \bR^n$.
For this purpose, choose an arbitrary element 
$W \oplus \tilde{W} \in \mfs(\mfk \oplus \mfk)$  
and decompose it as  
\be
W \oplus \tilde{W} 
= (A + \ri \kappa \bsone_N) 
\oplus (B - \ri \kappa \bsone_N),
\ee 
where $A, B \in \mfk$ are traceless matrices 
and $\kappa \in \bR$. Upon defining the 
anti-Hermitian tridiagonal matrix 
$T \in \bR^{n \times n}$ 
with the only non-zero entries 
\be
T_{a, a + 1} = - T_{a + 1, a}
= \sum_{c = 1}^a
\frac{A_{c, c} + B_{c, c}}{2 \ri g} \in \bR 
\qquad
(a \in \bN_{n - 1}),
\ee
we see that the $N \times N$ matrix 
$X = \diag(T, T)$ belongs to the subalgebra $\mfk$. 
Therefore $X$ generates a well-defined tangent 
vector $\delta \rho = [X, \xi(E)] \in T_{\xi(E)} \cO$.
Moreover, let $\delta y = 0$ and consider the
tangent vector 
$\delta Y = (\delta Y)_+ + (\delta Y)_- \in \mfg$ 
determined uniquely by the requirements 
\be
(\delta Y)_+ = - B + \ri \kappa \bsone_N 
= - \tilde{W}
\quad \mbox{and} \quad 
((\delta Y)_-)_\diag = 0,
\ee
together with 
\be
((\delta Y)_-)_\offdiag
= \sinh(\tilde{\ad}_Q)^{-1} 
(A_\offdiag - (\delta \rho)_\offdiag
- \cosh(\tilde{\ad}_Q) ((\delta Y)_+)_\offdiag) 
\in \mfa^\perp.
\ee
By applying the derivative operator on the 
distinguished tangent vector 
$\delta y \oplus \delta Y \oplus \delta \rho$
defined by the above conditions, a 
straightforward calculation based on 
(\ref{Jext_derivative}) yields
\be
(\ddd J^\ext)_x 
(\delta y \oplus \delta Y \oplus \delta \rho) 
= W \oplus \tilde{W}.
\ee
In other words, the derivative is onto.
\qedsymb

\smallskip
\noindent
\textbf{Corollary 4.}
\emph{
There is a unique smooth manifold structure on
$\mfL_0$ (\ref{cL_0}) such that the pair 
$(\mfL_0, \iota_0)$ with the natural inclusion 
$\iota_0 \colon \mfL_0 \hookrightarrow \cP^\ext$
is an embedded submanifold of $\cP^\ext$. Moreover,
the pair $(\cM^S, \Upsilon^S)$ provides an 
equivalent model for $(\mfL_0, \iota_0)$,
i.e. $\mfL_0 \cong \cM^S$.
}

\smallskip
\noindent
\textbf{Proof.}
As is known (see e.g. Proposition 1.6.18 in 
\cite{AM}), the previous lemma guarantees that 
the level set $\mfL_0 \subset \cP^\ext$ is an 
embedded submanifold of dimension
\be
\dim(\mfL_0) =
\dim(\cP^\ext) - \dim(\mfs(\mfk \oplus \mfk))
= 4 n^2 + 2 n - 1 = \dim(\cM^S).
\ee
From lemma 2 we also see that $\Upsilon^S$ is 
a smooth bijective immersion from $\cM^S$ onto 
$\mfL_0$, therefore they are necessarily 
diffeomorphic.
\qedsymb

Utilizing the model $(\cM^S, \Upsilon^S)$ of the 
level set $\mfL_0$, in the remaining part of this 
section we complete the reduction of 
$\cP^\ext$ at the zero value of the momentum map 
$J^\ext$. Note that the stabilizer subgroup of 
$0 \in \mfs(\mfk \oplus \mfk)$ is the full group 
$K \times K$, and the (residual) 
$K \times K$ action on $\cM^S$ takes the form
\be
(k_L, k_R) \acts \left(
q, p, (\eta_L, \eta_R) U(1)_*\right)
= \left( q, p, 
(k_L \eta_L, k_R \eta_R) U(1)_*\right),
\ee
for any $q \in \mfc$, $p \in \bR^n$ and
$k_L, k_R, \eta_L, \eta_R \in K$. Thus, it is 
obvious that the orbit space $\cM^S /(K \times K)$
can be identified with the base manifold of
the trivial fiber bundle
\be
\pi^S \colon 
\cM^S \twoheadrightarrow \cP^S,
\quad
\left( q, p, (\eta_L, \eta_R) U(1)_* \right)
\mapsto (q, p),
\label{pi}
\ee
therefore the reduced symplectic manifold 
can be identified as
\be
\cP^\ext /\!/_0 (K \times K) 
\cong \cM^S / (K \times K) 
\cong \cP^S.
\ee
Recall also that the reduced symplectic structure 
$\omega^S \in \Omega^2 (\cP^S)$ 
is uniquely determined by the condition
\be
(\pi^S) ^* \omega^S 
= (\Upsilon^S)^* \omega^\ext.
\label{omega_red_def}
\ee
Working out the above pull-backs, the 
reduced symplectic form can be determined
in a straightforward manner.

\smallskip
\noindent
\textbf{Theorem 5.}
\emph{
Up to a multiplicative constant, the globally
defined coordinate functions 
$q_a$, $p_a$ $(a \in \bN_n)$ 
provide a Darboux system on the reduced manifold 
$\cP^S$. More precisely, the reduced 
symplectic form can be written as $\omega^S 
= 2 \sum_{a = 1}^n \ddd q_a \wedge \ddd p_a$.
} 

\smallskip
\noindent
\textbf{Proof.}
Take an arbitrary 
$r = (q, p) \in \cP^S$
and let 
$x = (q, p, (\bsone_N, \bsone_N) U(1)_*) \in \cM^S$.
Observe that the point $x$ projects onto $r$, i.e. 
$\pi^S (x) = r$. Also, we have   
\be
(\ddd \pi^S)_x 
\frac{\partial}{\partial q_c} \bigg|_x 
= \frac{\partial}{\partial q_c} \bigg|_{r}
\quad \mbox{and} \quad
(\ddd \pi^S)_x 
\frac{\partial}{\partial p_c} \bigg|_x 
= \frac{\partial}{\partial p_c} \bigg|_{r},
\ee
for every $c \in \bN_n$. 
If 
$e_{k, l} \in \mfgl(N, \bC)$ $(k, l \in \bN_N)$ 
denotes the elementary matrix with entries 
$(e_{k, l})_{k', l'} 
= \delta_{k, k'} \delta_{l, l'}$, 
then by (\ref{derivative_Upsilon}) we can write
\be
(\ddd \Upsilon^S)_x 
\frac{\partial}{\partial q_c} \bigg|_x 
= e^Q W_c
\oplus [W_c, Z] \oplus 0
\quad \mbox{and} \quad
(\ddd \Upsilon^S)_x 
\frac{\partial}{\partial p_c} \bigg|_x 
= 0 \oplus W_c \oplus 0,
\ee 
where
\be 
W_c = e_{c, c} - e_{n + c, n + c} \in \mfa
\quad \mbox{and} \quad
Z = \cosh(\tilde{\ad}_Q) 
\sinh(\tilde{\ad}_Q)^{-2} \xi(E) \in \mfk.
\ee
Now, remembering the defining formulae 
of $\omega^\ext$ (\ref{omega_ext}), from 
(\ref{omega_red_def}) it is immediate that  
\be
\omega^S_{r} \left(
\frac{\partial}{\partial q_a} \bigg|_{r}, 
\frac{\partial}{\partial p_b} \bigg|_{r}
\right)
= \omega^\ext_{\Upsilon^S(x)} \left(
e^Q W_a \oplus [W_a, Z] \oplus 0, 
0 \oplus W_b \oplus 0 \right) 
= \langle W_a, W_b \rangle = 2 \delta_{a, b},
\ee
for any $a, b \in \bN_n$. The remaining components 
of $\omega^S$ come along the same line.
\qedsymb

\smallskip
\noindent
The above theorem is well-known in the
literature (see e.g. 
\cite{AvanBabelonTalon1994},
\cite{FeherPusztai2006}). 
After developing an analogous theorem for
the rational $C_n$ RSvD model in section 4, 
its importance will be transparent in
section 5.

\section{The phase space of the rational
$C_n$ RSvD model}
\setcounter{equation}{0}
Starting with this section we present our
new results on the $C_n$-type rational RSvD model
with two independent coupling parameters.
Our first goal is to derive the
natural phase space of the RSvD 
model from symplectic reduction. As the initial 
step of the reduction, we have to explore the 
properties of the level set $\mfL_0$ (\ref{cL_0}). 
Contrary to the Sutherland case, now we are 
solving the momentum map constraint 
(\ref{constraint}) by diagonalizing the Lie 
algebra component of the extended phase space 
(\ref{Pext}).

First, for each $a \in \bN_n$ let us consider
the complex-valued rational function 
\be
\mfc \ni \lambda 
= (\lambda_1, \ldots, \lambda_n)
\mapsto
z_a(\lambda) 
= - \left(1 + \frac{\ri g_2}
{\lambda_a} \right)
\prod_{\substack{d = 1 \\ (d \neq a)}}^n 
\left( 1 + \frac{2 \ri g}
{\lambda_a - \lambda_d} \right)
\left( 1 + \frac{2 \ri g}{
\lambda_a + \lambda_d} \right) \in \bC.
\label{z}
\ee
Remembering the manifold $\cP^R$ introduced 
in (\ref{cP_R}), we also define the 
matrix-valued function
\be
\cA \colon 
\cP^R \rightarrow \mfgl(N, \bC),
\quad
(\lambda, \theta) \mapsto \cA(\lambda, \theta),
\label{cA_def}
\ee
where the matrix entries lying in the diagonal
$n \times n$ blocks are given by the formulae
\begin{align}
& \cA_{a, b}(\lambda, \theta) 
= e^{\theta_a + \theta_b}
\vert z_a(\lambda) z_b(\lambda) \vert^\half 
\frac{2 \ri g}{2 \ri g + \lambda_a - \lambda_b},
\\
& \cA_{n + a, n + b}(\lambda, \theta) 
= e^{-\theta_a - \theta_b}
\frac{ \overline{z_a(\lambda)} z_b(\lambda)}
{\vert z_a(\lambda) z_b(\lambda) \vert^\half} 
\frac{2 \ri g}{2 \ri g - \lambda_a + \lambda_b},
\end{align}
meanwhile the matrix entries belonging to the
off-diagonal $n \times n$ blocks have the form
\be 
\cA_{a, n + b}(\lambda, \theta) 
= \overline{\cA_{n + b, a}(\lambda, \theta)} 
= e^{\theta_a - \theta_b} z_b(\lambda) 
\vert z_a(\lambda) z_b(\lambda)^{-1} \vert^\half
\frac{2 \ri g}{2 \ri g + \lambda_a + \lambda_b}
+ \frac{\ri(g - g_2)}{\ri g + \lambda_a} 
\delta_{a, b},
\label{cA_entries}
\ee
for any $a, b \in \bN_n$. In our paper 
\cite{Pusztai_JPA} we introduced (\ref{cA_def}) 
as a natural candidate for the Lax matrix of the 
rational $C_n$ RSvD model, and we are now ready
to prove this claim. As one of
the key objects of our discussion, in the following 
subsection we examine the main properties of $\cA$
in detail. The Cauchy matrices play a prominent role
in our analysis, so the reader may
find it useful to consult the appendix  
for the necessary background material.

\subsection{The properties of the Lax matrix $\cA$}
To facilitate the comparison with our earlier work 
on the structure of $\cA$, we borrow some 
notations from \cite{Pusztai_JPA}. 
First, for each $\lambda \in \bR^n$ let $x \in \bC^N$ 
denote the column vector with the purely imaginary 
components
\be
x_a = - x_{n + a} 
= \frac{\lambda_a}{2 \ri g} \in \ri \bR
\qquad 
(a \in \bN_n).
\label{x}
\ee
Upon introducing the real parameter
\be
\eps = 1 - g_2 g^{-1} \in \bR,
\label{eps}
\ee
the function $z_a$ (\ref{z}) can be easily
related to the rational function $w_a$ 
(\ref{C_type_w}) naturally appearing in 
the study of the standard Cauchy matrices 
(\ref{Cauchy_matrix}). In fact $z_a$
is a deformation of $w_a$, as can be seen
from the relationship
\be
z_a(\lambda) 
= - w_a(x) 
\left( 1 - \frac{\eps}{1 + 2 x_a} \right)
\qquad
(a \in \bN_n).
\label{z&w}
\ee
Next, for any 
$(\lambda, \theta) \in \cP^R$
we define the column vector 
$\cF(\lambda, \theta) \in \bC^N$
with components
\be 
\cF_a(\lambda, \theta) 
= e^{\theta_a} \vert z_a(\lambda) \vert^\half
\quad \mbox{and} \quad
\cF_{n + a}(\lambda, \theta) 
= e^{- \theta_a} \overline{z_a(\lambda)} 
\vert z_a(\lambda) \vert^{-\half},
\label{cF}
\ee
where $a \in \bN_n$. Then the entries of $\cA$
(\ref{cA_def}) can be succinctly written 
as
\be
\cA_{k, l} 
= \frac{ \cF_k \overline{\cF}_l 
+ \eps C_{k, l} }{1 + x_k - x_l}
\qquad
(k, l \in \bN_N). 
\label{cA}
\ee 
Giving a glance at (\ref{cA}), it is obvious 
that $\cA(\lambda, \theta)$ is a
Hermitian matrix. Our next goal is to show that
$\cA(\lambda, \theta)$ belongs to the Lie group 
$U(n, n)$ (\ref{G}). 

\smallskip
\noindent
\textbf{Proposition 6.}
\emph{
For every 
$(\lambda, \theta) \in \cP^R$
we have
$\cA(\lambda, \theta) C \cA(\lambda, \theta) 
= C$, i.e. $\cA(\lambda, \theta) \in U(n, n)$.
}

\smallskip
\noindent
\textbf{Proof.}
Pick an arbitrary 
$(\lambda, \theta) \in \cP^R$.
Let us introduce the column vector
$u(x) \in \bC^N$ with components
\be
u_k(x) = \frac{1}{1 + 2 x_k} 
\qquad
(k \in \bN_N),
\label{u}
\ee
and the $N \times N$ diagonal matrices
\be
D = \diag(\cF_1, \ldots, \cF_N)
\quad \mbox{and} \quad 
\cD(x) = \diag(u_1(x), \ldots, u_N(x)).
\ee
Note that $C u(x) = u(-x)$ and 
$C \cD(x) C = \cD(-x)$.
Recalling (\ref{cA}) we can write
\be
\cA = D \cC(x) D^* + \eps \cD(x) C,
\label{cA_cB_matrices}
\ee
where $\cC(x)$ stands for the standard 
$N \times N$ Cauchy matrix 
(\ref{Cauchy_matrix}) associated with 
the vector $x \in (\ri \bR)^N$ (\ref{x}). 
Thus it is clear that
\be
\begin{split}
\cA C \cA - C
= D ( &
\cC(x) D^* C D \cC(x) D^* C D
+ \eps \cC(x) C \cD(x) C D^* C D 
\\
& + \eps \cD(x) \cC(x) D^* C D
+ \eps^2 \cD(x)^2 - \bsone_N) D^{-1} C.
\end{split}
\label{ACA}
\ee
However, remembering (\ref{cF}), (\ref{z&w}) 
and (\ref{W}), we can write
\be
D^* C D 
= C \diag(z_1, \ldots, z_n, 
\overline{z}_1, \ldots, \overline{z}_n)
= C W(x) (\eps \cD(x) - \bsone_N).
\ee
Therefore, plugging this relationship 
into (\ref{ACA}) and applying 
(\ref{C_type_matrix_identities}), we obtain
\be
\cA C \cA - C
= \eps D (\bsone_N + \cC(x) C W(x)) 
\cD(x) (\bsone_N + \cC(x) C W(x)) 
(\eps \cD(x) - \bsone_N) D^{-1} C.
\ee
Now let us notice that
\be
\cC(x) C W(x) \cD(x) 
+ \cD(x) \cC(x) C W(x)
= 2 u(x) u(x)^* W(-x) C,
\ee
from where it is immediate that
\be
(\bsone_N + \cC(x) C W(x)) 
\cD(x) (\bsone_N + \cC(x) C W(x)) 
= 2 C \left( 
u(-x) + \cC(-x) W(x) u(x) 
\right) 
u(x)^* W(-x) C.
\ee
However, from (\ref{w_identity_2}) we see
that for each $k \in \bN_N$ we can write
\be
\left( 
u(-x) + \cC(-x) W(x) u(x) 
\right)_k 
= u_k(-x) 
+ \sum_{j = 1}^N \cC_{k, j}(- x) w_j(x) u_j(x)
= 0, 
\ee
which immediately leads to the conclusion
$\cA C \cA - C = 0$, i.e. $\cA C \cA = C$.
\qedsymb

To proceed further, let us note that the
image of $\mfp$ (\ref{mfkp}) under the 
exponential map can be identified with the 
positive definite elements of $U(n, n)$,
i.e. we have 
\be
\exp(\mfp) 
= \{ y \in U(n, n) \, | \, y > 0 \}.
\label{exp_mfp_characterization}
\ee
Keeping in mind this fact, the proof of the
following lemma is quite straightforward.

\smallskip
\noindent
\textbf{Lemma 7.}
\emph{
For every 
$(\lambda, \theta) \in \cP^R$
the Hermitian matrix $\cA(\lambda, \theta)$
belongs to $\exp(\mfp)$.
}

\smallskip
\noindent
\textbf{Proof.}
Take an arbitrary point
$(\lambda, \theta) \in \cP^R$
and keep it fixed. Note that when $\eps = 0$, 
the lemma is trivial. Indeed, in this special case 
$\cA$ (\ref{cA}) is a Cauchy-type matrix, thus a 
direct application of the determinant formula 
(\ref{Cauchy_determinant}) immediately shows 
that all its leading principal minors are 
positive. 

In order to verify the lemma for arbitrary values 
of the coupling parameters $g$ and $g_2$, let us 
notice that the dependence of $\cA$ (\ref{cA}) 
on the real parameter $\eps$ (\ref{eps}) is 
continuous. Moreover, according to proposition 6,
if we change this parameter 
continuously from $0$ to an arbitrary given value, 
then during the course of the deformation the 
matrix $\cA$ remains invertible and Hermitian. 
Therefore, by continuity, the eigenvalues cannot 
pass through zero, i.e., they remain positive. 
It is now evident
that for all $\eps$ the Hermitian matrix $\cA$ is 
a positive definite element of $U(n, n)$, whence by 
the identification (\ref{exp_mfp_characterization}) 
the proof is complete.
\qedsymb

An immediate consequence of the above 
lemma is that $\cA(\lambda, \theta)$ has a 
(unique) positive square root 
\be
\cR(\lambda, \theta) 
= \cA(\lambda, \theta)^\half,
\label{cR}
\ee
which also belongs to $\exp(\mfp)$. In particular,
the column vector
\be
\cV(\lambda, \theta) 
= \cA(\lambda, \theta)^{-\half} 
\cF(\lambda, \theta) 
\in \bC^N
\label{cV}
\ee
is well-defined for all 
$(\lambda, \theta) \in \cP^R$.  
The importance of the following algebraic
properties of $\cV$ will be clear in the next 
subsection.

\smallskip
\noindent
\textbf{Proposition 8.}
\emph{
For any $(\lambda, \theta) 
\in \cP^R$ we have
$\cV(\lambda, \theta)^* 
\cV(\lambda, \theta) = N$
and $C \cV(\lambda, \theta) 
+ \cV(\lambda, \theta) = 0$.
}

\smallskip
\noindent
\textbf{Proof.}
Recalling (\ref{z&w}), (\ref{cF}) and (\ref{cA}),
the definition of $\cV$ (\ref{cV}) leads to the
following expansion  
\be
\begin{split}
\cV^* \cV = \cF^* \cA^{-1} \cF 
= & \sum_{k, l = 1}^N 
\frac{w_k(-x) w_l(x)}{1 - x_k + x_l}
\\
& - \eps \left( \sum_{k, l = 1}^N 
\frac{w_k(-x) w_l(x)}{1 - x_k + x_l}
\left(\frac{1}{1 - 2 x_k} 
+ \frac{1}{1 + 2 x_l} \right)
+ \sum_{l = 1}^N \frac{w_l(x)}{1 + 2 x_l}
\right)
\\
& + \eps^2 \left(\sum_{k, l = 1}^N 
\frac{w_k(-x) w_l(x)}{1 - x_k + x_l}
\frac{1}{(1 - 2 x_k)(1 + 2 x_l)} 
+ \sum_{l = 1}^N \frac{w_l(x)}{(1 + 2 x_l)^2}
\right).
\end{split}
\ee
Due to (\ref{w_identity_1}) and (\ref{trW}), 
for the first sum we can write
\be
\sum_{k, l = 1}^N 
\frac{w_k(-x) w_l(x)}{1 - x_k + x_l}
= \sum_{k = 1}^N w_k(-x) 
\sum_{l = 1}^N \frac{w_l(x)}{1 + x_l - x_k}
= \sum_{k = 1}^N w_k(-x) = \tr(W(-x)) = N.
\ee
Making use of (\ref{w_identity_2}), the 
coefficient of $\eps$ can be rewritten as
\be
\sum_{k, l = 1}^N 
\frac{w_k(-x) w_l(x)}{1 - x_k + x_l}
\left(\frac{1}{1 - 2 x_k} 
+ \frac{1}{1 + 2 x_l} \right)
+ \sum_{l = 1}^N \frac{w_l(x)}{1 + 2 x_l} 
= 2 \sum_{k = 1}^N \frac{w_k(-x)}{1 - 2 x_k}
\sum_{l = 1}^N \frac{w_l(x)}{1 + 2 x_l} = 0.
\ee
Finally, using (\ref{w_identity_1}) and 
(\ref{w_identity_2}), for the coefficient 
of $\eps^2$ we obtain
\be
\begin{split}
\sum_{k, l = 1}^N 
& \frac{w_k(-x) w_l(x)}{1 - x_k + x_l}
\frac{1}{(1 - 2 x_k)(1 + 2 x_l)} 
+ \sum_{l = 1}^N 
\frac{w_l(x)}{(1 + 2 x_l)^2} =
\\
& = \sum_{l = 1}^N 
\frac{w_l(x)}{1 + 2 x_l}
\left( \sum_{k = 1}^N \frac{w_k(-x)}
{(1 - x_k + x_l)(1 - 2 x_k)}
+ \frac{1}{1 + 2 x_l} \right)
= 0. 
\end{split}
\ee
Putting the above formulae together
we end up with $\cV^* \cV = N$.

Next, notice that 
$C \cV + \cV = \cA^{-\half}(\cA C \cF + \cF)$.
However, recalling identities 
(\ref{w_identity_1}) and (\ref{w_identity_2}), 
for any $k \in \bN_N$ we can write
\be
(\cA C \cF)_k 
= \cF_k \left(
- \sum_{j = 1}^N 
\frac{w_j(-x)}{1 - x_j + x_k} 
+ \eps \sum_{j = 1}^N 
\frac{w_j(-x)}{(1 - x_j + x_k)(1 - 2 x_j)}
+ \eps \frac{1}{1 + 2 x_k} \right) 
= - \cF_k,
\label{cA_C_cF}
\ee
therefore the relationship $C \cV + \cV = 0$ 
also follows.
\qedsymb

\subsection{Parametrization of the level 
set $\mfL_0$}
At this point we are in a position to introduce 
an appropriate parametrization of the level set 
$\mfL_0$ (\ref{cL_0}) naturally induced by the 
diagonalization of the Lie algebra part of 
$\cP^\ext$ (\ref{Pext}). Indeed, take an arbitrary 
$\lambda 
= (\lambda_1, \ldots, \lambda_n) \in \bR^n$ 
and introduce the $N \times N$ diagonal matrix 
\be
\cL(\lambda) = \diag(\bslambda, -\bslambda)
= \diag(\lambda_1, \ldots, \lambda_n,
- \lambda_1, \ldots, - \lambda_n) \in \mfa,
\label{cL}
\ee
then the points of $\mfL_0$ can be 
characterized as follows.

\smallskip
\noindent
\textbf{Proposition 9.}
\emph{
For each $(y, Y, \rho) \in \mfL_0$ there exist
some $(\lambda, \theta) \in \cP^R$
and $\eta_L, \eta_R \in K$ such that
\be
y = \eta_L \cA(\lambda, \theta)^\half \eta_R^{-1},
\quad
Y = \eta_R \cL(\lambda) \eta_R^{-1},
\quad
\rho = \eta_L \xi(\cV(\lambda, \theta)) \eta_L^{-1}.
\label{proposition_9}  
\ee
}

\smallskip
\noindent
\textbf{Proof.}
Take an arbitrary $(y, Y, \rho) \in \mfL_0$.
Since $J^\ext(y, Y, \rho) = 0$, from 
(\ref{Jext}) we see that $Y_+ = 0$, i.e. 
$Y = Y_- \in \mfp$. Therefore, by 
(\ref{diagonalizing_mfp}), we can write
\be
Y = \eta_R \cL(\lambda) \eta_R^{-1}
\label{R_Y}
\ee 
with some $\eta_R \in K$ and 
$\lambda = (\lambda_1, \ldots, \lambda_n) 
\in \bR^n$
satisfying 
$\lambda_1 \geq \ldots \geq \lambda_n \geq 0$. 

Also, due to the global Cartan 
decomposition (\ref{global_Cartan_decomposition}), 
the group element $y$ can be uniquely
factorized as $y = y_- y_+$, where
$y_- \in \exp(\mfp)$ and $y_+ \in K$.
Upon introducing $\eta_L = y_+ \eta_R \in K$, 
let us observe that
$\eta_L^{-1} y \eta_R = \eta_L^{-1} y_- \eta_L 
\in \exp(\mfp)$, therefore 
\be
y = \eta_L e^\Lambda \eta_R^{-1}
\label{R_y}
\ee
with some $\Lambda \in \mfp$. 

Plugging (\ref{R_Y}) 
and (\ref{R_y}) into $J^\ext(y, Y, \rho) = 0$, 
and recalling (\ref{cO}), we also get
\be
0 = (y Y y)_+ + \rho 
= \eta_L \left( 
\sinh(\ad_\Lambda) \cL(\lambda) 
\right)\eta_L^{-1} + \xi(V),
\label{R_constraint_2}
\ee
where $\rho = \xi(V)$ with some complex column 
vector $V \in \bC^N$ satisfying $V^* V = N$ and 
$C V + V = 0$. Note that equation 
(\ref{R_constraint_2}) is equivalent to
\be
\sinh(\ad_\Lambda) \cL(\lambda) 
= - \xi(\eta_L^{-1} V),
\ee
which immediately translates into
\be
2 \ri g e^{2 \Lambda} + \cL(\lambda) e^{2 \Lambda}
- e^{2 \Lambda} \cL(\lambda)
= 2 \ri g (e^\Lambda \eta_L^{-1} V)
(e^\Lambda \eta_L^{-1} V)^* + 2 \ri (g - g_2) C.
\label{R_key_relation}
\ee
Fortunately the last two equations coincide with
the pivotal relations that permitted us to understand 
the scattering theory of the hyperbolic 
$C_n$ Sutherland model. Indeed, compare the above 
relations with equations (28) and (31) in 
\cite{Pusztai_JPA}. 
Notice also that these relations are the natural 
analogs of the commutation relations in Ruijsenaars' 
treatment on the action-angle duality between the 
hyperbolic Sutherland and the rational 
Ruijsenaars--Schneider models of type $A_n$ 
(see equation (2.4) in \cite{RuijCMP1988}).
Without repeating the arguments presented in section 
3 of \cite{Pusztai_JPA}, we simply quote the outcome 
of the analysis of equation (\ref{R_key_relation}). 
First, due to lemma 1 of \cite{Pusztai_JPA}, 
$\cL(\lambda)$ is a regular 
element\footnote{To be precise, in \cite{Pusztai_JPA}
this fact is proved under the mild assumption 
$g_2 \neq 2 g$. Because of the quadratic 
appearance of the coupling parameters in the 
Hamiltonian systems of our interest (see equations 
(\ref{H_Sutherland}) and (\ref{H_RSvD})), this 
technical condition does not mean any restriction 
on our results.}  
of $\mfp$, therefore 
$\lambda_1 > \ldots > \lambda_n > 0$,
i.e. $\lambda \in \mfc$. Second, by lemma 2 of 
\cite{Pusztai_JPA}, we can write
$e^{2 \Lambda} = \cA(\lambda, \theta)$
with some $\theta \in \bR^n$, whence by equation
(\ref{R_y}) we obtain
\be
y = \eta_L \cA(\lambda, \theta)^\half \eta_R^{-1}.
\label{y_OK}
\ee
Third, the analysis yields the relation
$e^\Lambda \eta_L^{-1} V = \cF(\lambda, \theta)$, 
too, therefore 
$V = \eta_L \cA(\lambda, \theta)^{-\half} 
\cF(\lambda, \theta)$,
i.e. 
\be
\rho = \xi(\eta_L \cV(\lambda, \theta)) 
= \eta_L \xi(\cV(\lambda, \theta)) \eta_L^{-1}.
\label{rho_OK}
\ee 
Having a look at on equations (\ref{R_Y}),
(\ref{y_OK}) and (\ref{rho_OK}), the proposition
follows.
\qedsymb

Obviously the above proposition is the complete
analog of proposition 1. Now let us introduce
the manifold
\be
\cM^R = \cP^R \times (K \times K) / U(1)_*.
\label{cM_R}
\ee
Developing the theory
parallel to the previous section, the following
lemma can be seen as the natural analog of
lemma 2. 

\smallskip
\noindent
\textbf{Lemma 10.}
\emph{
The map
\be
\Upsilon^R \colon \cM^R \rightarrow \mfL_0,
\quad
(\lambda, \theta, (\eta_L, \eta_R) U(1)_*)
\mapsto 
( \eta_L \cA(\lambda, \theta)^\half \eta_R^{-1},
\eta_R \cL(\lambda) \eta_R^{-1},
\eta_L \xi(\cV(\lambda, \theta)) \eta_L^{-1} )
\label{Upsilon_R}
\ee
is a diffeomorphism, i.e. the pair 
$(\cM^R, \Upsilon^R)$ provides an appropriate
equivalent model for the embedded submanifold 
$(\mfL_0, \iota_0)$. 
}

\smallskip
\noindent
\textbf{Proof.}
First, we check that $\Upsilon^R$ is a well-defined
map. For, take an arbitrary
$(\lambda, \theta, (\eta_L, \eta_R) U(1)_*) 
\in \cM^R$. As we have discussed after 
lemma 7, the square root of $\cA(\lambda, \theta)$
does belong to $\exp(\mfp)$, whence we have
\be
y 
= \eta_L \cA(\lambda, \theta)^\half \eta_R^{-1}
\in G.
\ee
Because of proposition 8, the column vector 
$\cV(\lambda, \theta)$ (\ref{cV}) generates a 
well-defined element of the orbit $\cO$ (\ref{cO}), 
therefore 
$\rho 
= \eta_L \xi(\cV(\lambda, \theta)) \eta_L^{-1} 
\in \cO$.
Also, note that 
$Y = \eta_R \cL(\lambda) \eta_R^{-1} \in \mfp$.
To proceed further we evaluate  $J^\ext$ 
(\ref{Jext}) at the point 
$(y, Y, \rho) \in \cP^\ext$ with the 
above components. Notice that
\be
(y Y y^{-1})_+ + \rho 
= \half \eta_L (
\cA(\lambda, \theta)^\half \cL(\lambda) 
\cA(\lambda, \theta)^{-\half}
- \cA(\lambda, \theta)^{-\half} \cL(\lambda) 
\cA(\lambda, \theta)^\half
+ 2 \xi(\cV(\lambda, \theta)) ) \eta_L^{-1}.
\ee
However, due to (\ref{xi}), (\ref{cV}) and 
(\ref{cA}), we can write
\be
\cA^\half \cL \cA^{-\half}
- \cA^{-\half} \cL \cA^\half
+ 2 \xi(\cV)
= - 2 \ri g \cA^{-\half} \left(
\cA + [(2 \ri g)^{-1} \cL, \cA] 
- \cF \cF^*
- \eps C \right) \cA^{-\half}
= 0,
\ee
therefore $(y Y y^{-1})_+ + \rho = 0$ 
immediately follows. Combining this fact with
the obvious relation $Y_+ = 0$, we see that 
$J^\ext(y, Y, \rho) = 0$, i.e. $\Upsilon^R$ 
is well-defined. 

Obviously $\Upsilon^R$ is a smooth map. 
Moreover, as we can infer from proposition 9, 
it is surjective. To complete the proof of 
the lemma, let us notice that $\Upsilon^R$ acts 
between manifolds of the same dimensions.
Thus it is enough to show that $\Upsilon^R$ 
is an injective immersion.

Let 
$(\lambda, \theta, (\eta_L, \eta_R) U(1)_*)$ 
and 
$(\tilde{\lambda}, \tilde{\theta}, 
(\tilde{\eta}_L, \tilde{\eta}_R) U(1)_*)$ 
be some points of $\cM^R$ and suppose that
\be
\Upsilon^R(\lambda, \theta, (
\eta_L, \eta_R) U(1)_*) 
= \Upsilon^R(\tilde{\lambda}, \tilde{\theta}, 
(\tilde{\eta}_L, \tilde{\eta}_R) U(1)_*).
\label{Upsilon_R_injective}
\ee
The $\mfg$-component of the above equation 
has the form 
\be
\eta_R \cL(\lambda) \eta_R^{-1}
= \tilde{\eta}_R \cL(\tilde{\lambda}) 
\tilde{\eta}_R^{-1}.
\ee
Since $\lambda, \tilde{\lambda} \in \mfc$, 
from (\ref{mfp_reg_parametrization}) we obtain
$\cL(\lambda) = \cL(\tilde{\lambda})$,
i.e. $\lambda = \tilde{\lambda}$. Moreover, 
we can write $\tilde{\eta}_R = \eta_R m$ 
with some unique group element 
$m = \diag(m_1, \ldots, m_N) \in M$
(\ref{M}).

Next, notice that the $G$-component of 
(\ref{Upsilon_R_injective}) yields 
\be
(\eta_L \cA(\lambda, \theta)^\half \eta_L^{-1}) 
( \eta_L \eta_R^{-1} ) 
= \eta_L \cA(\lambda, \theta)^\half \eta_R^{-1}
= \tilde{\eta}_L 
\cA(\tilde{\lambda}, \tilde{\theta})^\half 
\tilde{\eta}_R^{-1}
= (\tilde{\eta}_L 
\cA(\tilde{\lambda}, \tilde{\theta})^\half 
\tilde{\eta}_L^{-1})
( \tilde{\eta}_L \tilde{\eta}_R^{-1} ).
\ee
By the uniqueness of the global
Cartan decomposition 
(\ref{global_Cartan_decomposition}), we 
obtain $\tilde{\eta}_L = \eta_L m$ together
with the relationship
\be
\cA(\lambda, \theta)^\half 
= m \cA(\lambda, \tilde{\theta})^\half m^{-1}.
\ee
Now, by (\ref{cA}), the matrix equation 
$\cA(\lambda, \theta) 
= m \cA(\lambda, \tilde{\theta}) m^{-1}$
immediately leads to the relations
\be
\cF_k (\lambda, \theta) 
\overline{\cF_l (\lambda, \theta)}
= m_k \cF_k (\lambda, \tilde{\theta}) 
\overline{\cF_l (\lambda, \tilde{\theta})} 
m_l^{-1}
\qquad (k, l \in \bN_N).
\label{R_cF_relations}
\ee
Recalling the definition (\ref{cF}),
with $k = l = a$ $(a \in \bN_n)$ we get
$e^{2 \theta_a} = e^{2 \tilde{\theta}_a}$,
whence $\theta = \tilde{\theta}$. Plugging
this back into (\ref{R_cF_relations}),
for all $k, l \in \bN_N$ we obtain
$m_k = m_l$, therefore 
$m = e^{\ri \chi} \bsone_N$ 
with some $\chi \in \bR$.
It implies that 
$(\eta_L, \eta_R) U(1)_* 
= (\tilde{\eta}_L, \tilde{\eta}_R) U(1)_*$,
hence the verification of the injectivity
of $\Upsilon^R$ is complete.

Finally, we are showing that $\Upsilon^R$ 
is an immersion. For, let 
$x = 
(\lambda, \theta, (\eta_L, \eta_R) U(1)_*) 
\in \cM^R$ 
be an arbitrary point and take some tangent 
vector
\be
v 
= \delta \lambda \oplus \delta \theta \oplus 
(\eta_L, \eta_R) (X_L \oplus X_R) \mfu(1)_*
\in T_x \cM^R
\label{R_v}
\ee
generated by the tangent vectors 
$\delta \lambda \in T_\lambda \mfc \cong \bR^n$,
$\delta \theta \in T_\theta \bR^n \cong \bR^n$
and $X_L, X_R \in \mfk$.
Our first goal is to find the action of the
derivative 
\be
(\ddd \Upsilon^R)_x \colon 
T_x \cM^R \rightarrow T_{\Upsilon^R(x)} \mfL_0
\ee
on vector $v$. As in (\ref{cR}), let $\cR$ 
denote the square root of $\cA$. Also, let us 
introduce the shorthand notations 
$\delta \cL = \cL(\delta \lambda) \in \mfp$,
$\rho = \xi(\cV(\lambda, \theta)) \in \cO$,
together with the tangent vectors
\begin{align}
& \delta \cR 
= (\ddd \cR)_{(\lambda, \theta)} 
(\delta \lambda \oplus \delta \theta)
\in T_{\cR(\lambda, \theta)} G,  
\\ 
& \delta \cF = (\ddd \cF)_{(\lambda, \theta)} 
(\delta \lambda \oplus \delta \theta)
\in T_{\cF(\lambda, \theta)} \bC^N 
\cong \bC^N, 
\end{align}
and 
\be
\begin{split}
\delta \rho 
= & (\ddd (\xi \circ \cV))_{(\lambda, \theta)}
(\delta \lambda \oplus \delta \theta)
\\
= & \ri g \cR^{-1} \left( 
(\delta \cF) \cF^* + \cF (\delta \cF)^* 
- (\delta \cR) \cR^{-1} \cF \cF^*
- \cF \cF^* \cR^{-1} (\delta \cR) 
\right) \cR^{-1} \in T_\rho \cO.
\end{split}
\label{delta_rho}
\ee
Utilizing the above objects, one can easily 
verify that 
\be
(\ddd \Upsilon^R)_x v 
= \eta_L (\delta \cR + X_L \cR - \cR X_R) 
\eta_R^{-1}
\oplus \eta_R (\delta \cL - [\cL, X_R]) 
\eta_R^{-1} 
\oplus \eta_L ( \delta \rho - [\rho, X_L] ) 
\eta_L^{-1}.
\label{derivative_Upsilon_R}
\ee

To find the kernel of the linear transformation
$(\ddd \Upsilon^R)_x$, let us suppose that 
$v \in \ker((\ddd \Upsilon^R)_x)$. Note that both
$\cL$ and $\delta \cL$ are diagonal matrices,
meanwhile the commutator $[\cL, X_R]$ is 
off-diagonal. Therefore, the vanishing of the 
$\mfg$-component of (\ref{derivative_Upsilon_R}) 
entails the relations $\delta \cL = 0$ 
and $[\cL, X_R] = 0$. Whence 
$\delta \lambda = 0$, and by the regularity of $\cL$
the matrix $X_R$ must be diagonal, i.e. 
$X_R = \diag(\ri \bschi, \ri \bschi) \in \mfm$
with some $\chi \in \bR^n$. Combining this 
fact with the vanishing of the $G$-component 
of (\ref{derivative_Upsilon_R}), we obtain
\be 
X_L = \cR X_R \cR^{-1} - (\delta \cR) \cR^{-1}.
\label{R_XL_XR}
\ee
For simplicity, let us introduce the 
notation
\be
\delta \cA = (\ddd \cA)_{(\lambda, \theta)} 
(\delta \lambda \oplus \delta \theta)
\in T_{\cA(\lambda, \theta)} G. 
\label{delta_cA}
\ee 
Since $\cA = \cR^2$, the relationship between 
their derivatives leads to the formula 
\be
\delta \cA 
= (\delta \cR) \cR + \cR (\delta \cR). 
\label{deltacAcR}
\ee
Keeping in mind (\ref{delta_rho}), 
(\ref{R_XL_XR}) and (\ref{deltacAcR}), the 
vanishing of the $\cO$-component of the tangent
vector (\ref{derivative_Upsilon_R}) immediately 
yields
\be
(\delta \cF) \cF^* + \cF (\delta \cF)^*
- \cF \cF^* X_R + \cA X_R \cA^{-1} \cF \cF^*
- (\delta \cA) \cA^{-1} \cF \cF^* = 0.
\label{R_der_relations}
\ee
However, since $\delta \lambda = 0$, the 
derivative of the vector-valued function 
$\cF$ (\ref{cF}) comes easily. 
Namely, for the components of $\delta \cF$
we simply have 
\be
(\delta \cF)_a = (\delta \theta)_a \cF_a
\quad \mbox{and} \quad 
(\delta \cF)_{n + a} 
= - (\delta \theta)_a \cF_{n + a}
\qquad (a \in \bN_n).
\label{deltacF_components}
\ee
Upon introducing the $N \times N$ diagonal matrix
\be
D = \diag \left(
(\delta \theta)_1, \ldots, (\delta \theta)_n,
-(\delta \theta)_1, \ldots, -(\delta \theta)_n
\right) \in \mfa,
\label{delta_D}
\ee
the above relations (\ref{deltacF_components}) 
can be cast into the
simpler matrix form $\delta \cF = D \cF$. 
Moreover, by inspecting the derivative of 
$\cA$ (\ref{cA}), we obtain the concise 
expression
\be
(\delta \cA) \cA^{-1} 
= D + \cA D \cA^{-1}.
\ee
Plugging these formulae into 
(\ref{R_der_relations}), we get
\be
[X_R - D, \cA^{-1} \cF \cF^*] = 0.
\label{R_matrix_OK}
\ee
Remembering (\ref{cA_C_cF}), we can write 
$\cA C \cF = - \cF$, 
whence $\cA^{-1} \cF = - C \cF$. Therefore
the commutation relation (\ref{R_matrix_OK}) 
can be rewritten as
\be
(X_R - D)_{k, k} (C \cF)_k \overline{\cF}_l
= (C \cF)_k \overline{\cF}_l (X_R - D)_{l, l} 
\qquad
(\forall k, l \in \bN_N).
\ee
Since the components of $\cF$ are non-zero, 
it follows immediately that 
\be
(X_R - D)_{k, k} = (X_R - D)_{l, l}
\qquad
(\forall k, l \in \bN_N).
\label{R_components}
\ee
In particular, for each $a \in \bN_n$ we 
have
\be
\ri \chi_a - (\delta \theta)_a
= (X_R - D)_{a, a} = (X_R - D)_{n + a, n + a}
= \ri \chi_a + (\delta \theta)_a,
\ee
whence $(\delta \theta)_a = 0$, i.e. 
$\delta \theta = 0$. Plugging this 
back into (\ref{R_components}), we 
get $X_R = \ri \chi \bsone_N$ with some 
$\chi \in \bR$. Utilizing (\ref{R_XL_XR}), 
we obtain that $X_L = \ri \chi \bsone_N$, too, 
therefore $X_L \oplus X_R \in \mfu(1)_*$. 
That is, each component of the vector $v$ 
(\ref{R_v}) is zero, meaning that the kernel 
of the derivative $(\ddd \Upsilon^R)_x$ is 
trivial.
\qedsymb

\subsection{Canonical coordinates on the 
reduced phase space}
Now we are in a position to perform the
second, concluding step of the symplectic 
reduction. Without repeating our discussion 
in the paragraph preceding theorem 5, let us 
notice that the base manifold of the trivial 
fiber bundle 
\be
\pi^R \colon 
\cM^R \twoheadrightarrow \cP^R,
\quad
(\lambda, \theta, (\eta_L, \eta_R) U(1)_*) 
\rightarrow (\lambda, \theta) 
\label{pi_R}
\ee
provides an appropriate model for the
reduced phase space $\cP^\ext /\!/_0 (K \times K)$.
In order to obtain the reduced symplectic
structure, it is tempting to imitate the
proof of theorem 5. Since we had very 
explicit and simple formulae for the objects 
appearing in the definition of $\Upsilon^S$ 
(\ref{Upsilon_S}), in that case the computation 
of the pull-back $(\Upsilon^S)^* \omega^\ext$
was almost trivial, hence by (\ref{omega_red_def})
the reduced symplectic form was also immediate.
However, the definition of $\Upsilon^R$ 
(\ref{Upsilon_R}) involves the square root of 
$\cA$, for which we have no explicit formula.
Therefore a direct computation of the pull-back
$(\Upsilon^R)^* \omega^\ext$ seems to be hopeless.  
To circumvent this apparent difficulty, we instead
switch to the analysis of the reduced Poisson 
bracket $\{ \cdot \, , \cdot \}^R$ induced 
by the reduced symplectic form $\omega^R$ via 
a formula analogous to (\ref{PB}). 

In what follows, we introduce two families of 
auxiliary smooth functions defined on the extended 
phase space $\cP^\ext$ (\ref{Pext}). Namely, for each  
$r \in \bN$ we define 
\be
\varphi_r \colon \cP^\ext \rightarrow \bR,
\quad 
(y, Y, \rho) \mapsto \varphi_r(y, Y, \rho) 
= \frac{\tr(Y^r) + \tr((Y^*)^r)}{2 r}
\label{varphi}
\ee
and
\be
\Psi_r \colon \cP^\ext \rightarrow \bR,
\quad
(y, Y, \rho) \mapsto \Psi_r(y, Y, \rho)
= \frac{\tr(Y^r y^* Z(\rho) y)
+ \tr((Y^*)^r y^* Z(\rho) y)}{2},
\label{Psi}
\ee
where
$Z(\rho) 
= (\ri g)^{-1} \rho + \bsone_N - \eps C$ 
is a Hermitian $N \times N$ matrix. 
Take an arbitrary point 
$x = (y, Y, \rho) \in \cP^\ext$ and
an arbitrary tangent vector 
$\delta y \oplus \delta Y 
\oplus \delta \rho
\in T_x \cP^\ext$. Utilizing the
bilinear form (\ref{bilinear_form}),
for the derivative of $\varphi_r$ 
we can write 
\be
(\ddd \varphi_r)_x 
(\delta y \oplus \delta Y 
\oplus \delta \rho)
= \delta y \oplus \delta Y 
\oplus \delta \rho [ \varphi_r ]
= \begin{cases}
0, & \text{if $r$ is odd}, \\
\langle Y^{r - 1}, \delta Y \rangle, 
& \text{if $r$ is even}.
\end{cases}
\label{dvarphi}
\ee
A slightly longer computation also reveals
that the derivative of $\Psi_r$ has the
form
\be
\begin{split}
(\ddd \Psi_r)_x 
(\delta y \oplus \delta Y \oplus \delta \rho) 
= & \left \langle 
\frac{Y^r + (Y^*)^r}{2} y^* Z(\rho) y
- C y^* Z(\rho) y \frac{Y^r + (Y^*)^r}{2} C,
y^{-1} \delta y \right \rangle
\\
& + \left \langle \sum_{j = 0}^{r - 1}
\frac{
Y^{r - 1 - j} y^* Z(\rho) y Y^j
- C (Y^*)^{r - 1 - j} y^* Z(\rho) y (Y^*)^j C}
{2},
\delta Y \right\rangle
\\
& + \left \langle 
\frac{y (Y^r + (Y^*)^r) y^* + 
C y (Y^r + (Y^*)^r) y^* C}{4 \ri g}, 
\delta \rho \right \rangle.
\end{split}
\label{dPsi}
\ee
Recalling the definitions (\ref{omega}), 
(\ref{omega_cO}), (\ref{omega_ext}) and 
(\ref{X_H}), we can even determine the 
Hamiltonian vector fields
$\bsX_{\varphi_r} \in \mfX(\cP^\ext)$ and 
$\bsX_{\Psi_r} \in \mfX(\cP^\ext)$
generated by the functions $\varphi_r$ and 
$\Psi_r$, respectively. Indeed, at the point 
$x$ we have 
\be
(\bsX_{\varphi_r})_x 
= \begin{cases}
0, & \text{if $r$ is odd}, \\ 
y Y^{r - 1} \oplus 0 \oplus 0, 
& \text{if $r$ is even}.
\end{cases}
\label{X_varphi}
\ee
It is also immediate that we can write
\be
(\bsX_{\Psi_r})_x 
= \Delta y \oplus \Delta Y \oplus \Delta \rho
\in T_x \cP^\ext
\label{X_Psi}
\ee
with components
\begin{align}
\Delta y
& = \frac{y}{2} \sum_{j = 0}^{r - 1}
\left( Y^{r - 1 - j} y^* Z(\rho) y Y^j
- C (Y^*)^{r - 1 - j} y^* Z(\rho) y (Y^*)^j C
\right) \in T_y G,
\label{X_Psi_component1} \\
\Delta Y & = \half \left(
C (Y^*)^r y^* Z(\rho) y C +
C y^* Z(\rho) y Y^r C 
- (Y^*)^r y^* Z(\rho) y
- y^* Z(\rho) y Y^r \right) 
\in T_Y \mfg \cong \mfg,
\label{X_Psi_component2} \\
\Delta \rho & = \frac{1}{4 \ri g}
[y (Y^r + (Y^*)^r) y^* + 
C y (Y^r + (Y^*)^r) y^* C, \rho]
\in T_\rho \cO.
\label{X_Psi_component3}
\end{align}

To proceed further, we continue with some 
standard facts on the family of the 
$K \times K$-invariant smooth functions
\be
C^\infty_{K \times K}(\cP^\ext)
= \{ H \in C^\infty(\cP^\ext) \, | \,
H \circ \Phi^\ext_{(k_L, k_R)} = H 
\text{ for all } (k_L, k_R) \in K \times K \},
\ee
where $\Phi^\ext$ (\ref{Phi_ext}) stands for
the natural action of the Lie group 
$K \times K$ on the extended phase space 
$\cP^\ext$. 
As is well-known (see e.g. Theorem 4.3.5 in 
\cite{AM}), for each 
$H \in C^\infty_{K \times K}(\cP^\ext)$ there 
is a unique $H^R \in C^\infty(\cP^R)$ 
such that
\be
(\pi^R)^* H^R = (\Upsilon^R)^* H.
\label{reduced_Hamiltonian}
\ee
The above function $H^R$ is called the 
\emph{reduced Hamiltonian} associated with $H$.
In particular, the functions
$\varphi_r$ (\ref{varphi}) and $\Psi_r$ 
(\ref{Psi}) are obviously $K \times K$-invariant. 
Making use of (\ref{Upsilon_R}) and
(\ref{pi_R}), one can easily verify
that the corresponding
reduced Hamiltonians have the form
\be
\varphi_r^R = \begin{cases}
0, & \text{if $r$ is odd}, \\
\frac{2}{r} \sum_{a = 1}^n \lambda_a^r, 
& \text{if $r$ is even}, 
\end{cases}
\label{reduced_varphi}
\ee
and
\be
\Psi_r^R = \begin{cases}
2 \sum_{a = 1}^n 
\lambda_a^r \vert z_a \vert  
\sinh(2 \theta_a), 
& \text{if $r$ is odd}, \\
2 \sum_{a = 1}^n 
\lambda_a^r \vert z_a \vert  
\cosh(2 \theta_a),
& \text{if $r$ is even}. 
\end{cases}
\label{reduced_Psi}
\ee
From the perspective of symplectic reduction
it is a crucial fact that the family of the 
$K \times K$-invariant smooth functions forms 
a Poisson subalgebra of $C^\infty(\cP^\ext)$. 
Moreover, for any 
$H_1, H_2 \in C^\infty_{K \times K}(\cP^\ext)$
we have 
\be
(\pi^R)^* \{ H_1^R, H_2^R \}^R
= (\Upsilon^R)^* \{ H_1, H_2 \}^\ext.
\label{reduced_PB}
\ee
It is also immediate that the map 
\be
C^\infty_{K \times K}(\cP^\ext) \ni 
H \mapsto H^R 
\in C^\infty(\cP^R)
\ee 
is a Poisson algebra homomorphism. 
Based on 
the above observations, in a sequence of 
propositions we will show that the global 
coordinates 
$\lambda_a$ and $\theta_a$ $(a \in \bN_n)$ 
form a Darboux system on the reduced phase 
space $\cP^R$ (\ref{cP_R}).

\smallskip
\noindent
\textbf{Proposition 11.}
\emph{
For any $a, b \in \bN_n$ we have
$\{ \lambda_a, \lambda_b \}^R = 0$.
}

\smallskip
\noindent
\textbf{Proof.}
Let $r, s \in \bN$ be arbitrary 
\emph{even} numbers. Also, let 
$x = (y, Y, \rho) \in \cP^\ext$
be an arbitrary point; then
by (\ref{PB}), (\ref{dvarphi}) and 
(\ref{X_varphi}) we can write
\be
\{ \varphi_r, \varphi_s \}^\ext(x)
= (\bsX_{\varphi_s})_x [\varphi_r]
= y Y^{s - 1} \oplus 0 \oplus 0 [\varphi_r] 
= 0.
\ee
Due to (\ref{reduced_PB}) we obtain 
immediately that 
\be
\{ \varphi_r^R, \varphi_s^R \}^R = 0.
\label{varphivarphi1}
\ee
However, the above bracket can also be analyzed 
by exploiting the bilinearity, antisymmetry, 
Jacobi identity and Leibniz rule, characterizing 
any Poisson bracket. Indeed, recalling
(\ref{reduced_varphi}), we can write
\be
\{ \varphi_r^R, \varphi_s^R \}^R 
= \frac{4}{r s} \sum_{a, b = 1}^n 
\{ \lambda_a^r, \lambda_b^s \}^R
= 4 \sum_{a, b = 1}^n 
\lambda_a^{r - 1}
\{ \lambda_a, \lambda_b \}^R
\lambda_b^{s - 1}.
\label{varphivarphi2}
\ee
Now let us introduce the Vandermonde-type
$n \times n$ matrix $\mfV$ with entries
\be
\mfV_{a, b} = \lambda_a^{2 b - 1}
\qquad (a, b \in \bN_n).
\label{mfV}
\ee
Since $\lambda \in \mfc$, i.e. 
$\lambda_1 > \ldots > \lambda_n > 0$,
for the determinant of $\mfV$ we have
\be
\det(\mfV) 
= \begin{vmatrix}
\lambda_1 & \lambda_1^3 
& \ldots & \lambda_1^{2 n - 1} \\
\lambda_2 & \lambda_2^3 
& \ldots & \lambda_2^{2 n - 1} \\
\vdots & \vdots & \ddots & \vdots \\
\lambda_n & \lambda_n^3 
& \ldots & \lambda_n^{2 n - 1} \\
\end{vmatrix}
= \prod_{1 \leq c < d \leq n}
(\lambda_d^2 - \lambda_c^2)
\prod_{a = 1}^n \lambda_a 
\neq 0.
\label{det_mfV}
\ee
Also, we define the $n \times n$ matrix
$\Omega$ with entries 
$\Omega_{a, b} 
= \{ \lambda_a, \lambda_b \}^R$
$(a, b \in \bN_n)$.
Comparing equations (\ref{varphivarphi1}) and 
(\ref{varphivarphi2}), with the identifications 
$r = 2 c$ and $s = 2 d$ we can immediately 
write
\be
0 = \sum_{a, b = 1}^n 
\lambda_a^{2 c - 1} 
\{ \lambda_a, \lambda_b \}^R
\lambda_b^{2 d - 1} 
= \sum_{a, b = 1}^n 
\mfV_{a, c} \Omega_{a, b} \mfV_{b, d}
= \sum_{a, b = 1}^n 
(\mfV^*)_{c, a} \Omega_{a, b} \mfV_{b, d}
= (\mfV^* \Omega \mfV)_{c, d},
\ee
for all $c, d \in \bN_n$. These relations
can be cast into the matrix form 
$\mfV^* \Omega \mfV = 0$. However,
from equation (\ref{det_mfV}) we see 
that $\mfV$ is 
invertible, therefore $\Omega = 0$, i.e.
$\{ \lambda_a, \lambda_b \}^R = 0$
for all $a, b \in \bN_n$.
\qedsymb

\smallskip
\noindent
\textbf{Proposition 12.}
\emph{
For any $a, b \in \bN_n$ we have
$\{ \theta_a, \lambda_b \}^R 
= \half \delta_{a, b}$.
}

\smallskip
\noindent
\textbf{Proof.}
Take an \emph{odd} number $r$ and an 
\emph{even} number $s$. Let 
$x = (y, Y, \rho) \in \cP^\ext$ be 
an arbitrary point with the additional
assumption $Y^* = Y$. Recalling
(\ref{PB}), (\ref{dPsi}) and (\ref{X_varphi}), 
one can easily verify that 
\be
\{ \Psi_r, \varphi_s \}^\ext(x)
= (\bsX_{\varphi_s})_x [\Psi_r]
= y Y^{s - 1} \oplus 0 \oplus 0 [\Psi_r]
= 2 \Psi_{r + s - 1}(x).
\ee
Therefore, by (\ref{reduced_PB}), we obtain
the relationship
\be
\{ \Psi_r^R, \varphi_s^R \}^R
= 2 \Psi_{r + s - 1}^R.
\label{Psivarphi}
\ee
From equations (\ref{reduced_varphi}), 
(\ref{reduced_Psi}), and from the previous 
proposition it is immediate that the left 
hand side of the above equation can be 
rewritten as
\be
\{ \Psi_r^R, \varphi_s^R \}^R
= \frac{4}{s} \sum_{a, b = 1}^n 
\{ \lambda_a^r \vert z_a \vert
\sinh(2 \theta_a),
\lambda_b^s \}^R
= 4 \sum_{a = 1}^n 
\lambda_a^r \vert z_a \vert
\cosh(2 \theta_a) 
\sum_{b = 1}^n 
\{ 2 \theta_a, \lambda_b \}^R
\lambda_b^{s - 1}.
\ee
Comparing this expression with the 
right hand side of (\ref{Psivarphi}),
from (\ref{reduced_Psi}) we see that
\be
\sum_{a = 1}^n 
\lambda_a^r \vert z_a \vert
\cosh(2 \theta_a) \left( 
\sum_{b = 1}^n 
\{ 2 \theta_a, \lambda_b \}^R
\lambda_b^{s - 1} 
- \lambda_a^{s - 1} \right)
= 0.
\label{tl_key}
\ee
Upon introducing the $n \times n$ matrix
$\Xi$ with entries
\be
\Xi_{a, d} = \vert z_a \vert
\cosh(2 \theta_a) \left( 
\sum_{b = 1}^n 
\{ 2 \theta_a, \lambda_b \}^R
\lambda_b^{2 d - 1} 
- \lambda_a^{2 d - 1} \right)
\qquad 
(a, d \in \bN_n),
\label{Xi}
\ee
and recalling the Vandermonde-type matrix
$\mfV$ introduced in (\ref{mfV}), with the 
identifications $r = 2 c - 1$ and $s = 2 d$ 
the above equation (\ref{tl_key}) takes the 
form
\be
0 = \sum_{a = 1}^n 
\lambda_a^{2 c - 1} \Xi_{a, d}
= \sum_{a = 1}^n 
(\mfV^*)_{c, a} \Xi_{a, d}
= (\mfV^* \Xi)_{c, d},
\ee
for all $c, d \in \bN_n$. It simply means 
that $\mfV^* \Xi = 0$. However, due to 
(\ref{det_mfV}) the matrix $\mfV$ is invertible, 
hence $\Xi = 0$ also holds. Upon introducing
the $n \times n$ matrix $\Omega$ with entries
$\Omega_{a, b} 
= \{ 2 \theta_a, \lambda_b \}^R$
$(a, b \in \bN_n)$, the equation $\Xi_{a, d} = 0$
amounts to the requirement 
\be
0 = \sum_{b = 1}^n 
\{ 2 \theta_a, \lambda_b \}^R
\lambda_b^{2 d - 1} 
- \lambda_a^{2 d - 1} 
= \sum_{b = 1}^n 
\Omega_{a, b} \mfV_{b, d}
- \mfV_{a, d}
= (\Omega \mfV - \mfV)_{a, d},
\ee
for all $a, d \in \bN_n$. In 
matrix notation we can simply write 
$\Omega \mfV = \mfV$.
Again, since the Vandermonde-type
matrix $\mfV$ is invertible, it entails
$\Omega = \bsone_n$, i.e. 
$\{ 2 \theta_a, \lambda_b \}^R 
= \delta_{a, b}$ 
for all $a, b \in \bN_n$.
\qedsymb

\smallskip
\noindent
\textbf{Proposition 13.}
\emph{
For any $a, b \in \bN_n$ we have
$\{ \theta_a, \theta_b \}^R = 0$.
}

\smallskip
\noindent
\textbf{Proof.}
Though the proof of this proposition
is computationally more demanding,
the idea is the same as in the previous 
two propositions. Namely,
take an arbitrary pair of \emph{odd} numbers
$r, s \in \bN$ and compute the 
reduced Poisson bracket 
$\{ \Psi_r^R, \Psi_s^R \}^R$ 
in two different manners.

First, let
$(\lambda, \theta) \in \cP^R$
and define
$x = (\cA(\lambda, \theta)^\half , 
\cL(\lambda), \xi(\cV(\lambda, \theta))) 
\in \cP^\ext$.
Recalling (\ref{Upsilon_R}) and (\ref{pi_R}),
by (\ref{reduced_PB}) and (\ref{PB}) 
we can write
\be
\{ \Psi_r^R, \Psi_s^R \}^R 
(\lambda, \theta)
= \{ \Psi_r, \Psi_s \}^\ext (x)
= (\bsX_{\Psi_s})_x [\Psi_r] 
= (\ddd \Psi_r)_x (\bsX_{\Psi_s})_x.
\label{tt_1}
\ee
However, recalling (\ref{X_Psi}), we have  
\be
(\bsX_{\Psi_s})_x 
= \Delta y \oplus \Delta Y \oplus \Delta \rho,
\ee 
where the components can be determined from  
(\ref{X_Psi_component1}), (\ref{X_Psi_component2})
and (\ref{X_Psi_component3}), respectively. 
Now, by applying 
(\ref{dPsi}) on the components of the tangent
vector $(\bsX_{\Psi_s})_x$, we obtain
\be
\Delta y \oplus 0 \oplus 0 [\Psi_r]
= \sum_{a, b = 1}^n \sum_{j = 0}^{s - 1} 
\lambda_a^{r + j} \lambda_b^{s - 1 - j} 
\vert z_a(\lambda) \vert 
\vert z_b(\lambda) \vert 
(e^{2 \theta_a} - (-1)^j e^{-2 \theta_a})
(e^{2 \theta_b} + (-1)^j e^{-2 \theta_b})
\ee
and the similar expression
\be
0 \oplus \Delta Y \oplus 0 [\Psi_r]
= - \sum_{a, b = 1}^n \sum_{j = 0}^{r - 1} 
\lambda_a^{s + j} \lambda_b^{r - 1 - j} 
\vert z_a(\lambda) \vert 
\vert z_b(\lambda) \vert 
(e^{2 \theta_a} - (-1)^j e^{-2 \theta_a})
(e^{2 \theta_b} + (-1)^j e^{-2 \theta_b}),
\ee
together with the slightly more complicated 
derivative
\be
0 \oplus 0 \oplus \Delta \rho [\Psi_r]
= 4 \sum_{\substack{a, b = 1 \\ (a \neq b)}}^n
\lambda_a^r \lambda_b^s 
\vert z_a(\lambda) \vert 
\vert z_b(\lambda) \vert 
\left(
\frac{(\lambda_a - \lambda_b) 
\sinh(2 \theta_a + 2 \theta_b)}
{4 g^2 + (\lambda_a - \lambda_b)^2}
- \frac{(\lambda_a + \lambda_b) 
\sinh(2 \theta_a - 2 \theta_b)}
{4 g^2 + (\lambda_a + \lambda_b)^2}
\right).
\ee
Plugging these formulae into (\ref{tt_1}), 
simple algebraic manipulations lead to the
expression
\be
\begin{split}
\{ \Psi_r^R, \Psi_s^R \}^R
= & - 2 (r - s) \sum_{a = 1}^n 
\lambda_a^{r + s - 1} 
\vert z_a \vert^2 
\sinh(4 \theta_a)
\\
& + 16 g^2 
\sum_{\substack{a, b = 1 \\ (a \neq b)}}^n
\left(
\frac{\lambda_a^r \lambda_b^s 
\vert z_a \vert \vert z_b \vert
\sinh(2 \theta_a - 2 \theta_b)}
{(4 g^2 + (\lambda_a + \lambda_b)^2)
(\lambda_a + \lambda_b)}
- \frac{\lambda_a^r \lambda_b^s 
\vert z_a \vert \vert z_b \vert
\sinh(2 \theta_a + 2 \theta_b)}
{(4 g^2 + (\lambda_a - \lambda_b)^2)
(\lambda_a - \lambda_b)}
\right).
\end{split}
\label{tt_1_OK}
\ee

On the other hand, we can also work out
$\{ \Psi_r^R, \Psi_s^R \}^R$
relying only on the defining properties
(bilinearity, antisymmetry, etc.)
of the Poisson bracket. Indeed, utilizing
(\ref{reduced_Psi}) and the previous two 
propositions, one can easily verify that
\begin{align}
\{ \Psi_r^R, \Psi_s^R \}^R
= & 4 \sum_{a, b = 1}^n \left( 
\lambda_a^r \vert z_a \vert
\cosh(2 \theta_a) \sinh(2 \theta_b) 
\frac{\partial 
( \lambda_b^s \vert z_b \vert)}
{\partial \lambda_a}
- \lambda_b^s \vert z_b \vert
\sinh(2 \theta_a) \cosh(2 \theta_b) 
\frac{\partial 
( \lambda_a^r \vert z_a \vert)}
{\partial \lambda_b} \right)
\nonumber \\
& + 16 \sum_{a, b = 1}^n 
\lambda_a^r \lambda_b^s 
\vert z_a \vert \vert z_b \vert
\cosh(2 \theta_a) \cosh(2 \theta_b)
\{ \theta_a, \theta_b \}^R.
\label{tt_2}
\end{align}
At this point we need the partial  
derivatives of $\vert z_a \vert$ 
$(a \in \bN_n)$. However, recalling (\ref{z}), 
it is immediate that for any $a, b \in \bN_n$,
$a \neq b$, we have
\be
\frac{1}{\vert z_b \vert}
\frac{\partial \vert z_b \vert}
{\partial \lambda_a}
= \half 
\frac{\partial \ln (\vert z_b \vert^2)}
{\partial \lambda_a}
= - \frac{4 g^2}
{4 g^2 + (\lambda_a - \lambda_b)^2}
\frac{1}{\lambda_a - \lambda_b}
- \frac{4 g^2}
{4 g^2 + (\lambda_a + \lambda_b)^2}
\frac{1}{\lambda_a + \lambda_b}.
\ee
Along the same line we obtain
\be
\frac{1}{\vert z_a \vert}
\frac{\partial \vert z_a \vert}
{\partial \lambda_a}
= - \frac{g_2^2}{g_2^2 + \lambda_a^2} 
\frac{1}{\lambda_a}
- \sum_{\substack{b = 1 \\ (b \neq a)}}^n
\left( 
\frac{4 g^2}{4 g^2 
+ (\lambda_a - \lambda_b)^2}
\frac{1}{\lambda_a - \lambda_b}
+ \frac{4 g^2}{4 g^2 
+ (\lambda_a + \lambda_b)^2}
\frac{1}{\lambda_a + \lambda_b}
\right).
\ee
Now plugging these derivatives into (\ref{tt_2}),
we see at once that the resulting formula
coincides with the right hand side of 
(\ref{tt_1_OK}), up to some terms involving 
the desired Poisson brackets 
$\{ \theta_a, \theta_b \}^R$. 
More precisely, the comparison of (\ref{tt_2})
and (\ref{tt_1_OK}) yields
\be
\sum_{a, b = 1}^n 
\lambda_a^r \lambda_b^s 
\vert z_a \vert \vert z_b \vert
\cosh(2 \theta_a) \cosh(2 \theta_b)
\{ \theta_a, \theta_b \}^R = 0.
\label{tt_2_OK}
\ee
In order to infer 
$\{ \theta_a, \theta_b \}^R$
from the above equation, let us introduce 
the $n \times n$ matrix $\Omega$ with entries
\be
\Omega_{a, b} 
= \vert z_a \vert \vert z_b \vert
\cosh(2 \theta_a) \cosh(2 \theta_b)
\{ \theta_a, \theta_b \}^R
\qquad (a, b \in \bN_n).
\label{tt_Omega}
\ee
Recalling the Vandermode-type matrix $\mfV$ 
defined in (\ref{mfV}), with the identifications
$r = 2 c - 1$ and $s = 2 d - 1$ we can write 
\be
0 = \sum_{a, b = 1}^n 
\mfV_{a, c} \Omega_{a, b} \mfV_{b, d}
= \sum_{a, b = 1}^n 
(\mfV^*)_{c, a} \Omega_{a, b} \mfV_{b, d}
= (\mfV^* \Omega \mfV)_{c, d},
\ee
for all $c, d \in \bN_n$, i.e. we have the matrix
equation $\mfV^* \Omega \mfV = 0$. By 
the determinant formula (\ref{det_mfV}) the 
matrix $\mfV$ is invertible, 
whence $\Omega = 0$. Therefore, due to
(\ref{tt_Omega}), we end up with
$\{ \theta_a, \theta_b \}^R = 0$
for all $a, b \in \bN_n$.
\qedsymb

Putting together the contents of the previous 
three propositions we arrive at the most important 
technical result of the paper.

\smallskip
\noindent
\textbf{Theorem 14.}
\emph{
Up to an obvious rescaling, the globally defined
functions $\lambda_a$, $\theta_a$ $(a \in \bN_n)$ 
provide a canonical coordinate system on the 
reduced symplectic manifold $\cP^R$. More precisely, 
for the reduced symplectic form 
$\omega^R \in \Omega^2(\cP^R)$ we have
$\omega^R 
= 2 \sum_{a = 1}^n 
\ddd \theta_a \wedge \ddd \lambda_a$.
}

\smallskip
\noindent
Before turning to the consequences of the 
above theorem, we conclude this section with 
some remarks. First, recall that the existing 
proofs of the analogous theorem in the $A_n$ 
case are quite involved. They rely either on 
non-trivial analytic facts (see e.g. the proof 
of Theorem C1 in \cite{RuijCMP1988}), 
or on equally non-trivial functional identities 
(see the proof of Theorem 2 in 
\cite{FeherKlimcik0901}). 
On the other hand, the applied techniques in 
proving propositions 11, 12 and 13 are of purely 
algebraic nature. We believe that our proofs 
can easily be adapted to the $A_n$ root system,
providing an elementary and conceptually simpler 
understanding even in the $A_n$ case.
However, we find it a much more appealing
aspect of our approach that it can be naturally 
extended to understand the duality between the 
hyperbolic Sutherland and the rational RSvD models 
associated with the $BC_n$ root system. Since 
the $BC_n$ case is computationally even more 
demanding that the $C_n$ case, we wish to come 
back to the $BC_n$ problem in a later publication. 
 
\section{Action-angle duality and its consequences}
\setcounter{equation}{0}
Based on two different diagonalization
procedures, in the previous two sections 
we worked out the symplectic reduction
of the extended phase space 
$\cP^\ext = G \times \mfg \times \cO$ 
(\ref{Pext}) at the zero value of the 
momentum map $J^\ext$ (\ref{Jext}).
The derivation of the phase space of the
Sutherland models relies on the $KAK$ 
decomposition of the $G$-component of the 
level space $\mfL_0 = (J^\ext)^{-1}(\{ 0 \})$, 
meanwhile the phase space of the RSvD model 
comes from the diagonalization of the 
$\mfg$-component. Although the approaches are 
apparently different, still, the resulting 
symplectic manifolds $(\cP^S, \omega^S)$ and 
$(\cP^R, \omega^R)$ are two models of the 
\emph{same} reduced symplectic 
manifold $\cP^\ext /\!/_0 (K \times K)$. 
It is therefore immediate that there is a 
natural symplectomorphism 
$\cS \colon \cP^S \rightarrow \cP^R$
between $\cP^S$ and $\cP^R$. More precisely,
$\cS$ is a symplectomorphism making
the diagram
\be
\begin{split}
\xymatrix{
 & \cP^\ext & 
\\
\cM^S \ar[r]^{\Upsilon^S}_{\cong} 
\ar@{>>}[d]_{\pi^S} 
& \mfL_0 \ar@{^{(}->}[u]^{\iota_0} 
& \ar[l]_{\Upsilon^R}^{\cong} 
\ar@{>>}[d]^{\pi^R} \cM^R 
\\
\cP^S \ar[rr]_{\cS}^{\cong} & & \cP^R
}
\label{diagram}
\end{split}
\ee
commutative. 
Moreover, as we see in theorems 5 and 14, 
the standard coordinates 
$q_a$, $p_a$ $(a \in \bN_n)$ and
$\lambda_a$, $\theta_a$ $(a \in \bN_n)$
provide Darboux systems on $\cP^S$ and 
$\cP^R$, respectively. More precisely, 
they are canonical only up to some trivial 
rescaling. Nevertheless, by slightly abusing 
the terminology, in the rest of the section 
we shall simply refer to them as canonical 
coordinates. Now, for their pull-backs we 
introduce the shorthand notations
\be
\hat{\lambda}_a = \cS^* \lambda_a,
\quad
\hat{\theta}_a = \cS^* \theta_a,
\quad
\check{q}_a = (\cS^{-1})^* q_a,
\quad
\check{p}_a = (\cS^{-1})^* p_a,
\label{hats_and_checks}
\ee
for all $a \in \bN_n$.
Since $\cS$ is a symplectomorphism, it
is obvious that the
functions $\hat{\lambda}_a, 
\hat{\theta}_a \in C^\infty(\cP^S)$ 
$(a \in \bN_n)$ provide a new family of
(global) canonical coordinates on $\cP^S$, 
meanwhile the functions 
$\check{q}_a, \check{p}_a \in C^\infty(\cP^R)$
$(a \in \bN_n)$ are new canonical coordinates 
on $\cP^R$. At this point we are in a position 
to harvest some of the immediate consequences 
of the dual reduction picture (\ref{diagram}). 
  
Starting with the Sutherland side of the
dual reduction picture, take an arbitrary 
$(q, p) \in \cP^S$ and consider the point
\be
x 
= \Upsilon^S(q, p, (\bsone_N, \bsone_N) U(1)_*) 
= (e^Q, L(q, p), \xi(E)) \in \mfL_0.
\label{xS_1}
\ee
Making use of the parametrization $\Upsilon^R$
(\ref{Upsilon_R})
and introducing the shorthand notation
$(\lambda, \theta) = \cS(q, p) \in \cP^R$,
from (\ref{diagram}) it is immediate that
\be
x = (\eta_L \cA(\lambda, \theta)^\half \eta_R^{-1},
\eta_R \cL(\lambda) \eta_R^{-1},
\eta_L \xi(\cV(\lambda, \theta)) \eta_L^{-1}),
\label{xS_2}
\ee
with some $\eta_L, \eta_R \in K$. Comparing 
(\ref{xS_1}) and (\ref{xS_2}) it is clear that
\be
L(q, p) = \eta_R \cL(\lambda) \eta_R^{-1},
\label{L_and_cL}
\ee
therefore the spectrum of $L(q, p)$ can be 
identified as 
\be
\sigma(L(q, p)) 
= \{ \pm \hat{\lambda}_a (q, p) 
\, | \,
a \in \bN_n \}.
\label{sigmaL}
\ee
Now take an arbitrary real-valued $\Ad$-invariant 
smooth function $F \colon \mfg \rightarrow \bR$, 
i.e. we require that
\be
F(y Y y^{-1}) = F(Y)
\qquad
(\forall y \in G, \forall Y \in \mfg).
\label{invariant_F}
\ee
Recalling the notations introduced 
in (\ref{hats_and_checks}), from 
the relationship (\ref{L_and_cL}) 
we see that
\be
F \circ L 
= F( \cL(\hat{\lambda}_1,  
\ldots, \hat{\lambda}_n) ),
\ee
i.e. the composite function 
$F \circ L \in C^\infty(\cP^S)$
depends only on the functions 
$\hat{\lambda}_a$
$(a \in \bN_n)$. Therefore the 
canonical coordinates  $\hat{\lambda}_a$, 
$\hat{\theta}_a$ $(a \in \bN_n)$
provide an \emph{action-angle system}  
for the Hamiltonian system
$(\cP^S, \omega^S, F \circ L)$. 
It is worth noting that the action variables 
$\hat{\lambda}_a$ are actually the positive 
eigenvalues of $L$ (see (\ref{sigmaL})), meanwhile 
the angle coordinates are the pull-backs of 
the impulses $\theta_a$ of the RSvD picture. 
It is a trivial consequence of the above 
observations that during the evolution of the
dynamics governed by the Hamiltonian $F \circ L$, 
the Hermitian matrix $L$ (\ref{L}) undergoes an 
\emph{isospectral deformation}. Moreover, by
(\ref{sigmaL}), the positive eigenvalues of $L$ 
provide $n$ functionally independent smooth
functions in involution, therefore $L$ 
is indeed a Lax matrix for the system
$(\cP^S, \omega^S, F \circ L)$. 
Notice also that with the aid of the real-valued
$\Ad$-invariant function
\be
F_2(Y) = \frac{1}{4} \langle Y, Y \rangle
= \frac{1}{4} \tr(Y^2)
\qquad
(Y \in \mfg)
\label{F_2}
\ee
we can recover the hyperbolic $C_n$ Sutherland
system with two independent coupling constants. 
Indeed, one can easily verify that 
\be
F_2 \circ L = H^S_{C_n},
\ee 
therefore we have complete control over the
Lax matrix and the action-angle variables of 
the model (\ref{H_Sutherland}). 
We mention in passing that though 
the Lax matrix of the hyperbolic $C_n$ Sutherland 
model has been known for decades (see e.g. the 
survey \cite{OlshaPere}), 
to our knowledge the construction of action-angle 
coordinates for the non-$A_n$-type Sutherland 
models has not been carried out in the literature 
before.

Having finished our discussion on the Sutherland
picture, we now develop a parallel theory in the 
dual RSvD picture, too. For, take an arbitrary
$(\lambda, \theta) \in \cP^R$ and let
\be
x = (\cA(\lambda, \theta)^\half, \cL(\lambda), 
\xi(\cV(\lambda, \theta)) )\in \mfL_0.
\label{xR_1}
\ee
Recalling $\Upsilon^S$ (\ref{Upsilon_S}) and
defining 
$(q, p) = \cS^{-1}(\lambda, \theta) \in \cP^S$,
from (\ref{diagram}) it is evident that
\be
x = (\eta_L e^Q \eta_R^{-1}, 
\eta_R L(q, p) \eta_R^{-1}, 
\eta_L \xi(E) \eta_L^{-1})
\label{xR_2}
\ee
with some $\eta_L, \eta_R \in K$.
Comparing the $G$-components of (\ref{xR_1})
and (\ref{xR_2}), we obtain
\be
\cA(\lambda, \theta)^\half 
= \eta_L e^Q \eta_R^{-1},
\label{sqrtcA_and_eQ}
\ee
which in turn immediately yields
$\cA(\lambda, \theta) 
= \eta_L e^{2 Q} \eta_L^{-1}$.
Therefore the spectrum of the positive 
definite matrix $\cA(\lambda, \theta)$ has the 
form
\be
\sigma(\cA(\lambda, \theta)) 
= \{ 
e^{\pm 2 \check{q}_a(\lambda, \theta)}
\, | \,
a \in \bN_n \}.
\label{sigmacA}
\ee
Now take an arbitrary real-valued 
$K \times K$-invariant smooth function 
$f \colon G \rightarrow \bR$.
More precisely, we require the invariance of 
$f$ under the $K \times K$-action (\ref{KKonG}), 
i.e. we impose
\be
f( (k_L, k_R) \acts y) 
= f (k_L y k_R^{-1}) = f(y)
\qquad
(\forall (k_L, k_R) \in K \times K, 
\forall y \in G).
\label{invariant_f}
\ee
From (\ref{sqrtcA_and_eQ}) it is obvious
that
\be
f \circ \cA^\half
= f (\diag(
e^{\check{q}_1} , \ldots, e^{\check{q}_n},
e^{-\check{q}_1} , \ldots, e^{-\check{q}_n})),
\ee
i.e. the composite function 
$f \circ \cA^\half \in C^\infty(\cP^R)$
depends only on the coordinates
$\check{q}_a$ $(a \in \bN_n)$.
It is now immediate that the canonical
coordinates 
$\check{q}_a$, $\check{p}_a$ $(a \in \bN_n)$
give rise to an \emph{action-angle system} 
for the mechanical system 
$(\cP^R, \omega^R, f \circ \cA^\half)$. 
For purpose of interpretation we note
that the action variables $\check{q}_a$ are 
provided by the positive eigenvalues of 
$\ln(\cA) / 2$ (see (\ref{sigmacA})), meanwhile 
the angle variables come from the pull-backs 
of the canonical impulses $p_a$ appearing in
the Sutherland picture. Rephrasing the
above observations, we see that the positive
eigenvalues of $\ln(A) / 2$ form a commuting
family of $n$ functionally independent
smooth functions. Moreover, if the dynamics
is generated by the Hamiltonian $f \circ \cA^\half$
induced by some $K \times K$-invariant function 
$f$, then the matrix $\cA$ (\ref{cA_def}) 
undergoes an \emph{isospectral deformation}. 
Therefore it is fully justified to call $\cA$ 
a \emph{Lax matrix} for the Hamiltonian system 
$(\cP^R, \omega^R, f \circ \cA^\half)$.
To make contact with the RSvD models, let us
notice that the real-valued $K \times K$-invariant 
function
\be
f_1(y) = \half \tr(y y^*)
\qquad 
(y \in G)
\label{f_1}
\ee
generates the rational $C_n$ RSvD model 
(\ref{H_RSvD}) with two independent coupling 
parameters, i.e. we have
\be
f_1 \circ \cA^\half = H^R_{C_n}.
\ee 
To sum up, besides the Lax matrix $\cA$, from 
the dual reduction picture (\ref{diagram}) 
naturally emerges an appropriate action-angle 
system of canonical coordinates for the most 
general rational $C_n$ RSvD model. 
Both the Lax matrix $\cA$ and
the construction of the action-angle variables 
for the $C_n$ RSvD system appear to be new 
results.

Apart from a natural construction of 
action-angle coordinates, the symplectic reduction 
framework provides also nice solution algorithms 
for both the Sutherland and the RSvD models. 
Starting with the Sutherland picture, take
an arbitrary $\Ad$-invariant smooth function
$F \colon \mfg \rightarrow \bR$ (\ref{invariant_F}).
Let
\be
\pr_\mfg \colon \cP^\ext = G \times \mfg \times \cO
\twoheadrightarrow \mfg
\label{pr_mfg}
\ee
denote the natural projection onto the factor 
$\mfg$; then clearly 
$\pr_\mfg^* F = F \circ \pr_\mfg 
\in C^\infty_{K \times K}(\cP^\ext)$.
Notice that in the Sutherland picture the 
reduced Hamiltonian induced by the 
$K \times K$-invariant smooth function 
$\pr_\mfg^* F$ is exactly $F \circ L$. 
Thus, as is known from the theory of the 
symplectic reductions 
(see e.g. Theorem 4.3.5 in \cite{AM}), 
the Hamiltonian flows of $\pr_\mfg^* F$ flowing 
on the level space 
$(J^\ext)^{-1}( \{ 0 \} ) = \mfL_0 \cong \cM^S$ 
project onto the Hamiltonian flows of the
reduced Hamiltonian system 
$(\cP^S, \omega^S, F \circ L)$. To understand
the Hamiltonian flows of $\pr_\mfg^* F$ we 
need the $\mfg$-valued \emph{gradient} $\nabla F$ 
of $F$, which is defined by the formula
\be
(\ddd F)_Y (\delta Y)
= \langle \nabla F(Y), \delta Y \rangle
\qquad
(Y \in \mfg, \delta Y \in T_Y \mfg \cong \mfg).
\label{grad_F}
\ee
Let us notice that the 
infinitesimal version of the $\Ad$-invariance 
of $F$ takes the form
\be
[ Y, \nabla F(Y) ] = 0
\qquad (\forall Y \in \mfg),
\ee
whence from equations (\ref{X_H}) and 
(\ref{omega_ext}) 
we see that at each point 
$(y, Y, \rho) \in \cP^\ext$ the Hamiltonian 
vector field 
$ \bsX_{\pr_\mfg^* F} \in \mfX(\cP^\ext)$ 
has the form
\be
(\bsX_{\pr_\mfg^* F})_{(y, Y, \rho)}
= ( y \nabla F(Y) )_y \oplus 0_Y \oplus 0_\rho
\in T_{(y, Y, \rho)} \cP^\ext.
\ee
It is now immediate that the Hamiltonian
flows of $\pr_\mfg^* F$ are \emph{complete}
and they have the form
\be
\bR \ni t 
\mapsto (y_0 e^{t \nabla F(Y_0)}, Y_0, \rho_0)
\in \cP^\ext,
\ee
with some $(y_0, Y_0, \rho_0) \in \cP^\ext$.
Therefore the flows generated by the reduced 
Hamiltonian $F \circ L$ are also complete. 
Moreover, if 
\be
\bR \ni t \mapsto (q(t), p(t)) \in \cP^S
\label{reduced_flow_S}
\ee
is an arbitrary flow of the reduced Hamiltonian 
$F \circ L$, and if we introduce the shorthand
notation $L_0 = L(q(0), p(0))$,
then the flow of $\pr_\mfg^* F$ passing through
the point 
$(e^{Q(0)}, L_0, \xi(E)) \in \mfL_0$
projects onto the flow (\ref{reduced_flow_S}). 
More precisely, for each $t \in \bR$ there 
is some $(\eta_L(t), \eta_R(t)) \in K \times K$ 
such that
\be
(e^{Q(0)} e^{t \nabla F(L_0)}, 
L_0, \xi(E))
= (\eta_L(t) e^{Q(t)} \eta_R(t)^{-1},
\eta_R(t) L(q(t), p(t)) \eta_R(t)^{-1},
\eta_L(t) \xi(E) \eta_L(t)^{-1}).
\label{flow_S}
\ee
Let us observe that the $\mfg$-component
of the above equation has the form
\be
L_0 
= \eta_R(t) L(q(t), p(t)) \eta_R(t)^{-1},
\ee 
which confirms the fact that during the time 
evolution of the dynamics the Lax matrix $L$ 
(\ref{L}) does undergo an isospectral deformation.
More interestingly, from the $G$-component 
of (\ref{flow_S}) it is clear that
\be
e^{Q(0)} e^{t \nabla F(L_0)} 
e^{t \nabla F(L_0)^*} e^{Q(0)}
= \eta_L(t) e^{2 Q(t)} \eta_L(t)^{-1},
\ee
which immediately yields the spectral 
identification
\be
\sigma (e^{2 Q(t)}) 
= \sigma (e^{2 Q(0)} e^{t \nabla F(L_0)} 
e^{t \nabla F(L_0)^*}).
\ee
Our conclusion is that the diagonal matrix $Q(t)$, 
and so the trajectory $q(t) \in \mfc$ $(t \in \bR)$, 
can easily be recovered simply by diagonalizing 
the matrix flow 
\be
t \mapsto e^{2 Q(0)} e^{t \nabla F(L_0)} 
e^{t \nabla F(L_0)^*}.
\ee 
In particular, by choosing the quadratic 
function $F = F_2$ (\ref{F_2}), we can reconstruct 
the flows of the hyperbolic $C_n$ Sutherland 
model (\ref{H_Sutherland}) by diagonalizing
the exponential matrix flow
\be
t \mapsto e^{2 Q(0)} e^{t L_0}.
\label{Sutherland_matrix_flow}
\ee
Here we used the obvious fact that 
$\nabla F_2 (L_0) = L_0 / 2 \in \mfp$. 

Turning to the RSvD side of the dual reduction 
picture (\ref{diagram}), take an arbitrary
$K \times K$-invariant smooth function 
$f \colon G \rightarrow \bR$ (\ref{invariant_f})
defined on the group $G$. If
\be
\pr_G \colon \cP^\ext = G \times \mfg \times \cO
\twoheadrightarrow G
\label{pr_G}
\ee
denotes the natural projection onto $G$, then 
clearly 
$\pr_G^* f = f \circ \pr_G 
\in C^\infty_{K \times K}(\cP^\ext)$, whence 
$\pr_G^* f$ survives the reduction. More 
precisely, in the Ruijsenaars picture 
the corresponding reduced 
Hamiltonian turns out to be $f \circ \cA^\half$. 
Therefore the Hamiltonian flows of $\pr_G^* f$ 
staying on the level space 
$(J^\ext)^{-1}(\{ 0 \}) = \mfL_0 \cong \cM^R$
project onto the Hamiltonian flows of the 
reduced system 
$(\cP^R, \omega^R, f \circ \cA^\half)$.
To ease the calculations, let us introduce the 
$\mfg$-valued \emph{gradient} $\nabla f$ of $f$ 
by the requirement
\be
(\ddd f)_y (\delta y) 
= \langle \nabla f(y), y^{-1} \delta y \rangle
\qquad
(y \in G, \delta y \in T_y G).
\label{grad_f}
\ee
Utilizing the gradient, 
for the Hamiltonian vector field 
$\bsX_{\pr_G^* f} \in \mfX(\cP^\ext)$ 
at each point $(y, Y, \rho) \in \cP^\ext$ we can 
write
\be
(\bsX_{\pr_G^* f})_{(y, Y, \rho)}
= 0_y \oplus (- \nabla f(y) )_Y \oplus 0_\rho
\in T_{(y, Y, \rho)} \cP^\ext.
\ee
Clearly the flows generated by the Hamiltonian
$\pr_G^* f$ are \emph{complete}, having the form
\be
\bR \ni t \mapsto 
(y_0, Y_0 - t \nabla f(y_0), \rho_0) \in \cP^\ext, 
\label{R_unreduced_flow}
\ee
with some $(y_0, Y_0, \rho_0) \in \cP^\ext$.
Therefore the flows of the reduced Hamiltonian
$f \circ \cA^\half$ are also complete. 
Now take an arbitrary Hamiltonian flow 
\be
\bR \ni t 
\mapsto (\lambda(t), \theta(t)) \in \cP^R
\label{R_flow}
\ee 
of the reduced system 
$(\cP^R, \omega^R, f \circ \cA^\half)$.
By realizing (\ref{R_flow}) as a projection
of an appropriate flow (\ref{R_unreduced_flow})
on the levels space $\mfL_0 \cong \cM^R$, it
is clear that for each $t \in \bR$ we have
\begin{multline}
(\cA_0^\half, 
\cL_0 - t \nabla f(\cA_0^\half),
\xi(\cV_0)) \\
= (\eta_L(t) \cA(\lambda(t), \theta(t))^\half 
\eta_R(t)^{-1},
\eta_R(t) \cL(\lambda(t)) \eta_R(t)^{-1},
\eta_L(t) \xi(\cV(\lambda(t), \theta(t))) 
\eta_L(t)^{-1})
\label{flow_R}
\end{multline}
with some $\eta_L(t), \eta_R(t) \in K$, where
$\cA_0 = \cA(\lambda(0), \theta(0))$,
$\cL_0 = \cL(\lambda(0))$ and
$\cV_0 = \cV(\lambda(0), \theta(0))$. By 
inspecting the $G$-component of (\ref{flow_R})
it is immediate that
\be
\cA_0 
= \eta_L(t) \cA(\lambda(t), \theta(t)) 
\eta_L(t)^{-1},
\ee
which clearly shows that the evolution of the
Lax matrix $\cA$ (\ref{cA_def}) is isospectral. 
From our perspective it is more interesting that
the $\mfg$-component of (\ref{flow_R}) has the
form
\be
\cL_0 - t \nabla f (\cA_0^\half)
= \eta_R(t) \cL(\lambda(t)) \eta_R(t)^{-1},
\ee
from where we obtain the spectral identification
\be
\sigma (\cL(\lambda(t))) 
= \sigma(\cL_0 - t \nabla f (\cA_0^\half)).
\ee
The outcome of our analysis is that the
diagonal matrix $\cL(\lambda(t))$, and so
$\lambda(t) \in \mfc$ $(t \in \bR)$, can
be determined by diagonalizing the 
\emph{linear} matrix flow
\be
t \mapsto \cL_0 - t \nabla f (\cA_0^\half).
\label{RSvD_matrix_flow}
\ee
In particular, with the choice $f = f_1$ 
(\ref{f_1}) the above diagonalizing procedure 
permits us to construct the trajectories of 
the most general rational $C_n$ RSvD model 
(\ref{H_RSvD}).

Based on the action-angle duality and the 
above simple solution 
algorithms one can build up the scattering 
theory for both the Sutherland and the RSvD 
models. In this respect the main thrust  
comes from the intimate relationship 
between the scattering properties of the 
Calogero--Moser--Sutherland-type many-particle 
systems and the scattering characteristics 
of certain integrable soliton equations. 
The connection between the $A_n$-type
particle systems and the soliton equations
defined on the whole line is well understood
(see e.g. \cite{RuijSchneider}, 
\cite{RuijFiniteDimSolitonSystems} and 
\cite{BabelonBernard}),
but the link between the non-$A_n$-type particle 
systems and the soliton systems defined on the 
half-line is far less elaborated (see e.g. 
\cite{KapustinSkorik}). Motivated by this
intriguing relationship, the study of  
the temporal asymptotics of the exponential 
matrix flow (\ref{Sutherland_matrix_flow}) 
in our recent paper \cite{Pusztai_JPA} led 
us to the conclusion that the hyperbolic 
$C_n$ Sutherland model is a 
\emph{pure soliton system} with a 
\emph{factorized} scattering map. In the same 
spirit, by exploring the temporal asymptotics 
of the matrix flow (\ref{RSvD_matrix_flow}), 
one can work out the scattering map of the 
rational $C_n$ RSvD model as well. However,
a decent analysis would require the study
of the M\o ller wave transformations, too,
so to keep the paper in a reasonable length
we defer the scattering theoretic analysis
into a later publication.

To conclude this section, let us notice that 
apart from its interest in the theory of 
integrable many-particle systems and soliton 
equations, our results on the action-angle 
duality of the $C_n$-type models may have 
applications in the theory of random matrices 
as well. Built on the Lax matrices of 
the  Calogero--Moser--Sutherland-type many-particle 
systems associated with the $A_n$ root system,
the authors of the recent papers 
\cite{Bogomolny_PRL} and \cite{Bogomolny} 
have constructed certain 
\emph{integrable random matrix ensembles} 
characterized by spectral statistical properties 
`intermediate' between the Poisson and the 
GUE statistics. It turns out that in their 
investigation the main technical tool is the 
action-angle duality between the  
Calogero--Moser--Sutherland and the 
Ruijsenaars--Schneider integrable models of 
type $A_n$. Utilizing the Lax matrices $L$ 
(\ref{L}), $\cA$ (\ref{cA_def}), and the dual 
reduction picture (\ref{diagram}), it appears 
to be a challenging exercise to extend their
considerations onto root systems other 
that $A_n$. 

\renewcommand{\thesection}{A}
\section{Cauchy matrices and functional identities}
\renewcommand{\theequation}{A.\arabic{equation}}
\setcounter{equation}{0}
To understand the symplectic aspects of the 
rational $C_n$ RSvD model, especially the 
properties of its Lax matrix (\ref{cA_def}), 
we need some functional identities involving 
certain rational functions closely related to 
the theory of Cauchy matrices. The following 
identity, that we have learned from the appendix 
of \cite{van_Diejen_Duke_2000}, is of great 
importance in the study of both the Cauchy
matrices and the RSvD models. 
 
\smallskip
\noindent
\textbf{Lemma A1.} 
\emph{
Let $M, N \in \bN$ be arbitrary positive integers 
satisfying $M \leq N$, and take two families of 
arbitrary complex numbers 
$\alpha_1, \ldots, \alpha_M$ and 
$\beta_1, \ldots, \beta_N$ with 
$\beta_k \neq \beta_l$ for $k \neq l$. 
Then for any complex number
$z \in \bC 
\setminus \{ \beta_1, \ldots, \beta_N \}$
we have
\be
\frac{\prod_{k = 1}^M (z - \alpha_k)}
{\prod_{l = 1}^N (z - \beta_l)} 
= \delta_{M, N} 
+ \sum_{j = 1}^N \frac{1}{z - \beta_j} 
\frac{\prod_{k = 1}^M (\beta_j - \alpha_k)}
{\prod_{\substack{ l = 1 \\(l \neq j)}}^N 
(\beta_j - \beta_l)}. 
\label{key_identity}
\ee
} 

\smallskip
\noindent
Notice that the above identity can be 
verified by simple residue calculus.

Now, with any complex $N$-tuple 
$x = (x_1, \ldots, x_N) \in \bC^N$
we associate the standard $N \times N$ Cauchy 
matrix $\cC(x)$ with entries
\be
\cC_{k, l} (x) 
= \frac{1}{1 + x_k - x_l}
\qquad
(k, l \in \bN_N).
\label{Cauchy_matrix} 
\ee
The characteristic properties of the Cauchy
matrices can be found in any advanced textbook
on linear algebra (see e.g. \cite{Prasolov}).
Recall that the determinant of $\cC(x)$ is 
given by
\be
\det(\cC(x))
= \prod_{1 \leq k < l \leq N} 
\frac{1}{1 - (x_k - x_l)^{-2}}.
\label{Cauchy_determinant}
\ee
In order to get a concise formula for 
the inverse of the Cauchy matrix $\cC(x)$, 
for each $j \in \bN_N$ we define the rational 
function
\be
w_j(x) 
= \prod_{\substack{k = 1 \\ (k \neq j)}}^N
\frac{1 + x_j - x_k}{x_j - x_k}. 
\label{w}
\ee
A direct application of (\ref{key_identity})
immediately leads to the identity
\be
\sum_{j = 1}^N \frac{w_j(x)}{1 + x_j - x_k} = 1
\qquad (\forall k \in \bN_N).
\label{w_identity_1}
\ee
Now, let us introduce the $N \times N$ diagonal
matrix 
\be
W(x) 
= \diag(w_1(x), \ldots, w_N(x)).
\label{W}
\ee
Utilizing (\ref{key_identity}), one can 
easily verify e.g. by induction on $N$ that
\be 
\tr(W(x)) = N.
\label{trW}
\ee 
Most importantly, with the aid of the diagonal
matrix (\ref{W}) filled in with the rational 
functions (\ref{w}), the inverse of the Cauchy
matrix $\cC(x)$ can be computed by the formula   
\be
\cC(x)^{-1} = W(-x) \cC(-x) W(x).
\label{Cauchy_inverse}
\ee

By specializing the above identities,
we can produce some particularly useful 
relations suitable for analyzing the Lax
matrix of the $C_n$ RSvD model. 

\smallskip
\noindent
\textbf{Proposition A2.}
\emph{
Let $n \in \bN$, $N = 2 n$, and take an 
arbitrary complex vector 
$x = (x_1, \ldots, x_N) \in \bC^N$ satisfying
\be
x_{n + a} = - x_a
\qquad (\forall a \in \bN_n).
\label{x_assumption}
\ee
Under this assumption the rational functions 
(\ref{w}) can be rewritten as 
\be
w_a(x) = w_{n + a}(-x) 
= \left( 1 + \frac{1}{2 x_a} \right)
\prod_{\substack{d = 1 \\ (d \neq a)}}^n
\left( 1 + \frac{1}{x_a - x_d} \right)
\left( 1 + \frac{1}{x_a + x_d} \right)
\qquad (a \in \bN_n),
\label{C_type_w}
\ee
and for each $k \in \bN_N$ they enjoy the 
functional identities
\be
\sum_{j = 1}^N \frac{w_j(x)}{1 + 2 x_j} = 0
\quad \mbox{and} \quad
\sum_{j = 1}^N 
\frac{w_j(x)}{(1 + x_j - x_k)(1 + 2 x_j)}
= - \frac{1}{1 - 2 x_k}.
\label{w_identity_2}
\ee
Moreover, the diagonal matrix $W(x)$ (\ref{W}) 
and the Cauchy matrix $\cC(x)$ 
(\ref{Cauchy_matrix}) generated by the complex 
vector $x$ (\ref{x_assumption}) satisfy the 
relationships
\be 
C W(x) C = W(-x),
\quad
C \cC(x) C = \cC(-x),
\quad
\left( \cC(x) C W(x) \right)^2 = \bsone_N.
\label{C_type_matrix_identities}
\ee
}

\smallskip
\noindent
\textbf{Proof.}
We prove only the identities displayed in
(\ref{w_identity_2}). With the 
identifications 
$M = N$, $z = - 1 / 2$, $\alpha_l = x_l - 1$,
$\beta_l = x_l$ $(l \in \bN_N)$,
the application of the identity 
(\ref{key_identity}) immediately yields 
\be
\sum_{j = 1}^N \frac{w_j(x)}{1 + 2 x_j}
= - \half \sum_{j = 1}^N
\frac{1}{(-\half) - x_j}
\prod_{\substack{k = 1 \\ (k \neq j)}}^N 
\frac{x_j - (x_k - 1)}{x_j - x_k}
= \half \left(
1 - \prod_{c = 1}^n 
\frac{(x_c - \half)(x_{n + c} - \half)}
{(x_c + \half)(x_{n + c} + \half)} 
\right)
= 0.
\label{w_C}
\ee
Notice that the assumption (\ref{x_assumption})
on $x$ is crucial in the last step
of the algebraic manipulations.
Finally, take an arbitrary number $k \in \bN_N$; 
then from (\ref{w_C}) and (\ref{w_identity_1}) 
it follows that
\be
\sum_{j = 1}^N
\frac{w_j(x)}{(1 + x_j - x_k)(1 + 2 x_j)}
= \frac{1}{1 - 2 x_k} \left(
2 \sum_{j = 1}^N \frac{w_j(x)}{1 + 2 x_j}
- \sum_{j = 1}^N \frac{w_j(x)}{1 + x_j - x_k}
\right) = - \frac{1}{1 - 2 x_k}.
\ee
The rest of the proposition is trivial.
\qedsymb

\medskip
\noindent
\textbf{Acknowledgments.}
This work was partially supported by the Hungarian
Scientific Research Fund (OTKA) under grant
K 77400.


\end{document}